\documentclass[11pt]{article}
\usepackage{makecell}
\usepackage{verbatim,amsmath,amssymb}
\usepackage{xcolor}
\usepackage{epsfig,float,color}
\usepackage{epstopdf}
\usepackage{geometry}
\usepackage{setspace}
\usepackage{wrapfig}
\usepackage{hyperref}
\usepackage[utf8]{inputenc}
\usepackage{graphicx}
\usepackage{flushend}
\usepackage[linesnumbered,ruled]{algorithm2e}

\usepackage[font={footnotesize}]{caption}
\usepackage[font={footnotesize}]{subcaption}
\usepackage{footnote}
\usepackage{multicol}
\usepackage{multirow}
\usepackage{enumitem}   
\usepackage{soul}
\usepackage{framed}
\usepackage[titletoc,title]{appendix}
\usepackage{bm}
\usepackage{siunitx}
\usepackage{cite}
\usepackage{tikz}
\usepackage{pgfplots}
\usetikzlibrary{shapes.geometric, arrows.meta, positioning}
\tikzstyle{process} = [
rectangle,
rounded corners,
minimum width=4.8cm,
minimum height=1.2cm,
align=center,
text centered,
draw=black,
fill=cyan!40
]
\tikzstyle{arrow} = [thick, ->, >=Stealth]
\usepackage{pgfplots}
\pgfplotsset{compat=1.14} 
\usepgfplotslibrary{colormaps}
\usepackage[makeroom]{cancel} 
\usepackage{hyperref}

\renewcommand{\bar}[1]{\tilde{#1}}

\newcolumntype{C}[1]{>{\centering\arraybackslash}m{#1}}

\definecolor{darkblue}{rgb}{0,0,1}
\hypersetup{pdftex=true, colorlinks=true, breaklinks=true, linkcolor=darkblue, menucolor=darkblue, pagecolor=darkblue, citecolor=darkblue, urlcolor=darkblue}

\begin{document}
	
	\begin{center}
         \Large{\bf{Topology optimization with buckling constraints: Adaptive eigenvalue aggregation and modality identification}}
	\end{center}

    \begin{center}
		Badvelu Pranay Prabha$^{*,\dagger,+}$, Prabhat Kumar$^{\dagger,}$$\footnote{Corresponding author: pkumar@mae.iith.ac.in}$
		
		\vspace{4mm}
        \small{$*$\textit{School of Mechanical Sciences, Indian Institute of Technology Bhubaneswar, Odisha 752050, India}}\\
        \small{$+$\textit{Department of Mechanical Engineering, Indian Institute of Science, Bengaluru, Karnataka 560012, India}}\\
          \small{$\dagger$\textit{Department of Mechanical and Aerospace Engineering, Indian Institute of Technology Hyderabad, Kandi, Telangana 502285, India}}
	\end{center}

	\vspace{3mm}
	\rule{\linewidth}{.15mm}
   
{\bf Abstract:}
This paper presents novel approaches to tackle two key challenges in topology optimization problems with buckling constraints: determining how many eigenvalues to aggregate and identifying the eigenvalue modality at the optimal design. We demonstrate the mathematical inconsistencies that arise when choosing an arbitrary fixed number of eigenvalues for aggregation. Specifically, if the multiplicity of the critical eigenvalues is smaller than the chosen subset size, coalescence with higher-order eigenvalues can occur. Neglecting these during aggregation causes incorrect sensitivities, while simply increasing the preselected eigenvalue count unnecessarily raises computational costs. To resolve this issue, we propose an approach that dynamically determines the exact number of eigenvalues required at each optimization iteration. The proposed approach reduces computation time significantly while maintaining competitive performance relative to the conventional method across the numerical experiments presented.
	
Additionally, accurately predicting modality at the optimal point is crucial for understanding the buckled mode shapes of optimized designs. A new optimization formulation is proposed using the difference between eigenvalues to capture the true eigenvalue modality. This formulation is first validated on a classical 1D clamped column, successfully predicting both the bimodal solution and the optimal eigenvectors. The approach demonstrates that the optimal solutions for the wall reinforcement and shear-loaded hinged plate problems for the given set of parameters are trimodal and tetramodal, respectively.

	{\textbf {Keywords:}  Aggregation functions; Multimodal eigenvalues; Fr\'echet differentiability; Buckling constraints, Topology optimization}

	\vspace{-4mm}
	\rule{\linewidth}{.15mm}

\section{Introduction}
	Early investigations into optimization problems aimed at maximizing buckling load under volume constraints typically neglected the effects of multimodality. Olhoff and Rasmussen~\cite{olhoff1977single} first demonstrated that the optimal solution to a clamped-clamped column is inherently bimodal. Decades later, Seyranian et al. \cite{seyranian1994multiple} provided a comprehensive overview of the optimality conditions for general multimodal eigenvalue problems in their review paper. Utilizing the foundational optimality conditions provided in~\cite{seyranian1994multiple}, Neves et al. \cite{neves1995generalized} determined the optimal topology of 2D continuum structures under buckling constraints. Gravesen et al. \cite{gravesen2011sensitivities} established that symmetric polynomials of repeated eigenvalues are Fr\'echet differentiable, even when the individual eigenvalues themselves are not differentiable. Gao and Ma \cite{gao2015topology} applied symmetric polynomials to effectively impose buckling constraints within the topology optimization (TO) framework.

    TO is a computational design technique that determines an optimal material layout for a given problem while optimizing the intended objective under the given geometrical/physical constraints. The demand for TO-based approaches is continually rising, as they provide efficient, robust, and optimized design solutions for the problems at hand~\cite {sigmund2013topology,deaton2014survey}. To date, many methods have been proposed for TO; see~\cite{bendsoe1988generating,sigmund2013topology,deaton2014survey,allaire2004structural,xie1993simple,singh2025normalized}. Herein, we confine ourselves to the method involving buckling constraints with aggregation functions within a density-based TO framework. In this framework, the design domain is parameterized using standard/advanced finite elements (FEs)~\cite{kumar2023honeytop90}, and each element is assigned a design variable that ranges from 0 to 1. Design variables with 0 indicate a void region, whereas those with 1 represent an actual material area.

Aggregation functions have long been used in the optimization community to model non-smooth functions. The two most popular functions are the Kreisselmeier–Steinhauser (KS) function, proposed in 1980 \cite{kreisselmeier1980systematic}, and the $p$-norm function, introduced by Park and Kikuchi in 1995 \cite{park1995extensions}. Expanding beyond these conventional ones, Kennedy and Hicken~\cite{kennedy2015improved} introduced a new class of aggregation functions known as induced aggregates, thoroughly examined their properties, and their numerical implementation. They also proposed a post-optimality technique to mitigate the adverse effects of a finite aggregation parameter on the optimal solution. The literature on aggregation functions has been further enriched by several subsequent studies, including work by Kennedy on constraint strategies~\cite{kennedy2015strategies} and adjoint sensitivity~\cite{kennedy2016adjoint}, as well as a comprehensive evaluation of these methods by Lambe et al.~\cite{lambe2017evaluation}.
    
Initially, aggregation functions were widely adopted to solve stress-constrained optimization problems, in which numerous local stress constraints are reformulated into a single global constraint. For a comprehensive overview of techniques for handling stress-constrained problems, the reader can refer to several studies in the literature~\cite{paris2009topology,paris2010block,luo2013enhanced,verbart2017unified,da2021local,gao2015improved}.
 
  The earliest application of aggregation functions to eigenvalue problems was introduced by Chen et al.~\cite{46e63d1c814e4ce6bd6c8ba3010734b4}. They employed smoothing functions, similar to the KS function, to approximate the maximum-eigenvalue function. Torri et al.~\cite{torii2015modeling} used the $p$-norm function to impose global stability constraints on truss-like structures modeled with frame elements. Torii and Faria~\cite{torii2017structural} further investigated the mathematical properties of the $p$-norm function, deriving detailed sensitivity expressions to maximize the natural frequencies of vibrating continuum structures. Subsequently, Chin and Kennedy~\cite{chin2016large}, Dienemann et al.~\cite{dienemann2018considering}, and Leader et al.~\cite{doi:10.2514/1.J057777} applied these methods on large-scale high-resolution 3D TO problems with buckling, stress, and natural frequency constraints, respectively, thereby demonstrating the robustness and scalability of the aggregation function.

    Ferrari and Sigmund~\cite{ferrari2019revisiting} provided the intricacies of TO under buckling constraints. They specifically highlighted the utility of aggregation functions for addressing buckling-related challenges. Building on this work, Ferrari et al.~\cite{ferrari2021topology} introduced a compact 250-line MATLAB code capable of handling compliance, volume, and buckling loads as either objectives or constraints by employing the KS-aggregation function. Recently, Nishioka et al.~\cite{nishioka2025minimization} investigated the minimization of the maximum generalized eigenvalue, and comprehensively discussed the mathematical properties of the optimization problem using the calculus of non-differentiable functions. They subsequently proposed smoothing methods to numerically solve the problem, along with convergence studies and heuristic techniques to reduce computational cost. 
    
	This paper addresses two important issues with aggregation functions in structural optimization that, to the best of the authors' knowledge, remain unaddressed in the literature: (1) automating the selection of the number of eigenvalues to be aggregated rather than predefining, and (2) finding the eigenvalue modality at the optimum. Conventional techniques rely on a predefined number of eigenvalues in the aggregation step. When aggregating  $m$ (predefined) eigenvalues, the $m$-th and the next eigenvalues may coalesce during a TO iteration. Determining derivatives using $m$-eigenvalues can yield incorrect results if an arbitrary eigenvector basis is used. This issue becomes less pronounced when a larger $m$ is selected. However, choosing a high value for $m$ significantly increases the overall computational cost. To overcome these limitations, we introduce an adaptive scheme that dynamically selects only the required eigenvalues at each iteration. This approach maintains gradient accuracy during transitions, reduces overall computation time, and is more competitive than the traditional approach. The efficacy, success, and versatility of the proposed approach are demonstrated on three benchmark problems: the axially compressed column, wall reinforcement, and shear-loaded hinge problems. 
    
    For the second issue, standard aggregation methods cannot reliably identify the eigenvalue modality at the optimum and distinguish true multimodality from false coalescence.  This paper presents a novel formulation to determine the modality at the optimum by posing an unconstrained optimization problem. It primarily minimizes the difference between the eigenvalues while satisfying all constraints, while remaining close to the original optimal point. Finding modality at the optimum is important because it allows us to uncover all possible buckling modes by evaluating their linear combinations and gain complete physical insight into the buckled mode shapes of optimized designs. Using the proposed approach, we validate that the well-known fixed-fixed 1D compressed column problem is bimodal. Additionally, we find that the wall-reinforcement and shear-loaded-hinge problems are trimodal and tetramodal, respectively.

 The paper is organized as follows. Sec.~\ref{Sec2:Aggrefun&chall} defines the optimization problem and introduces the aggregation functions and their sensitivities. Sec.~\ref{Sec3:ChoiceEigen} uses an example to show how choosing an arbitrary number of eigenvalues causes mathematical errors, then proposes the $\epsilon$-ln threshold difference scheme to select the number of eigenvalues dynamically at each iteration. Three example problems: a compressed column, wall reinforcement, and a shear-Loaded hinged plate are solved to demonstrate that this dynamic strategy cuts computational time while matching the accuracy and convergence path of the traditional fixed-number method. Sec~\ref{Sec4:ModatOpt} first explores how to determine eigenvalue modality at the optimum point using a 1D column as an example. The section introduces a new strategy to predict eigenvalue coalescence (Sec.~\ref{SubSec4}), showing that the optimal designs for the wall-reinforcement and shear-loaded plate problems are trimodal and tetramodal, respectively. Finally, conclusions are drawn in Sec.~\ref{Sec5:Conc}.

\section{Aggregation functions and challenges}\label{Sec2:Aggrefun&chall}
	
The eigenvalue equation governing the critical buckling load factor of the discretized domain is written as~\cite{ferrari2021topology}:
	\begin{align}\label{Eq:MainbalanceEq}
		\mathbf{K}(\bm{x})\bm{\phi}_i(\bm{x}) = -\lambda_i(\bm{x})\mathbf{G}(\bm{x})\bm{\phi}_i(\bm{x})
	\end{align}
	where $\mathbf{K}$ and  $\mathbf{G}$ are the global stiffness and stress stiffness matrices, respectively. $\lambda_i$ represents the $i^{th}$ eigenvalue and $\phi_i$ indicates corresponding eigenvector. $N\times N$ provides the size of matrices $\mathbf{K}$ and $\mathbf{G}$, wherein $N$ indicates the total degrees of freedom of the parameterized design domain. Readers may refer to de Borst et al. (2012) \cite{de2012nonlinear} for a detailed FEM implementation of the buckling load equation. 
	
 One writes the optimization problems associated with buckling behavior typically in two forms: (i) buckling load as the objective function 
\begin{align}\label{Eq:Opt1}
	\max_{\boldsymbol{x}} \; \min_{i=1,\,\ldots\,,N} \lambda_i(\boldsymbol{x})
	\;\equiv\;
	\min_{\boldsymbol{x}} \; \max_{i=1,\,\ldots\,,N} \frac{1}{\lambda_i(\boldsymbol{x})}
	\quad
	\text{s.t.} \quad
	g(\bm{x}) \le 0,
\end{align}
and (ii) buckling load as a constraint 
\begin{align}\label{Eq:Opt2}
	\min_{\boldsymbol{x}} \quad & f(\boldsymbol{x}) \nonumber\\
	\text{s.t.} \quad 
	& \overline{\lambda} - \min_{i=1,\,\ldots\,,N} \lambda_i(\boldsymbol{x}) \le 0 .
\end{align}
Where $\bm{x}$ indicates the design vector, $\lambda_i(\boldsymbol{x})$ denotes the $i^{\text{th}}$ eigenvalue of the buckling problem, $f(\boldsymbol{x})$ represents the objective function (e.g., volume or compliance), $g(\boldsymbol{x})$ denotes the prescribed physical/geometrical constraint, and $\overline{\lambda}$ is the specified lower bound on the buckling load factor.

Both the optimization problems (Eq.~\ref{Eq:Opt1} and Eq.~\ref{Eq:Opt2}) require the gradient of $\min(\lambda_i)$ to be computed in each optimization iteration. However, the function $\min(\lambda_i)$ is usually not Fr\'echet differentiable~\cite{masur1979non,mach1995masur,haug1980design}. Therefore, traditional techniques/solvers fail to yield the optimal point~\cite{tadjbakhsh1962strongest}. This non-differentiability/non-smoothness occurs at the point of multimodality. One common approach to addressing multimodality is to use aggregation functions~\cite{ferrari2019revisiting,verbart2017unified}. Two of the most well-known aggregation functions are the Kreisselmeier-Steinhauser (KS) aggregation function \cite{kreisselmeier1980systematic} and the $p-$norm function \cite{park1995extensions}, wherein the level of approximation is governed by the aggregation parameter $\rho$ and $p$, respectively~\cite{kennedy2015improved}. One can use these aggregation functions to smooth $C=\min(\lambda_i)^{-1} = \max(\frac{1}{\lambda_i}) = \max(r_i)=r_{\max}$ for buckling optimization problems that ensure the existence of a classical Fr\'echet derivative~\cite{torii2017structural}, where $r_i = \frac{1}{\lambda_i}$. Mathematically, $\max(r_i)$ is  approximated using these functions as:
	\begin{equation}\label{Eq:KS_function}
	C_{\mathrm{KS}}(r_i, \rho)=\frac{1}{\rho} \ln \left( \sum_{i=1}^{m} \exp\left( \rho \, r_i \right) \right)=
r_{\max}+\frac{1}{\rho} \ln \left( \sum_{i=1}^{m} \exp\left( \rho \left( r_i - r_{\max} \right) \right) \right)
	\end{equation}
	\begin{equation}\label{Eq:P_function}
		C_\text{p}\left(r_{i}, p\right)=\left(\sum_{i=1}^{m}\left(r_{i}\right)^{p}\right)^{\frac{1}{p}}=r_{\max}\left(\sum_{i=1}^{m}\left(\frac{r_{i}}{r_{\max}}\right)^{p}\right)^{\frac{1}{p}} 
	\end{equation}
  where $C_\text{KS}\left(r_{i}, \rho\right)$ and $C_\text{p}\left(r_{i}, p\right)$ denote the KS-aggregation and $p$-norm approximations to $r_{\max}$, respectively. $m$ indicates the number of eigenvalues (buckling modes) included in the aggregation function.  Both forms of aggregation functions with and without $r_{max}$ in Eq.~\ref{Eq:KS_function} and Eq.~\ref{Eq:P_function} are mathematically equivalent. The latter expressions with $r_{max}$ ensure no numerical overflow during computer implementation~\cite{kennedy2015improved}. Typically, the KS-aggregation function is more preferred because it offers superior numerical stability~\cite{ferrari2019revisiting}.
In order to use a gradient-based optimization technique to solve either Eq.~\ref{Eq:Opt1} or Eq.~\ref{Eq:Opt2}, the gradients of $C_\text{KS}\left(r_i, \rho\right)$~\cite{ferrari2021topology} and $C_\text{p}\left(r_i, p\right)$ with respect to design variables are needed, which are determined as
\begin{equation}\label{Eq:KS_der}
\nabla C_{\mathrm{KS}}(r_i, \rho)
=
\frac{\displaystyle \sum_{i=1}^{m}
\exp\!\left( \rho (r_i - r_{\max}) \right)\, \partial r_i}
{\displaystyle \sum_{i=1}^{m}
\exp\!\left( \rho (r_i - r_{\max}) \right)}
=
\frac{\displaystyle \sum_{i=1}^{m}
\exp\!\left( \rho (r_i - r_{\max}) \right)
\left( -\, \bm{\phi}_i^{\top}
\left( \nabla \mathbf{G} + r_i \, \nabla \mathbf{K} \right)
\bm{\phi}_i \right)}
{\displaystyle \sum_{i=1}^{m}
\exp\!\left( \rho (r_i - r_{\max}) \right)}
\end{equation}
	
	\begin{equation}\label{Eq:pnorm_der}
		\nabla C_\mathrm{p}\left(r_i, p\right)
		= \frac{\displaystyle\sum_{i=1}^m\left(\frac{r_i}{r_{\max}}\right)^{p-1} \partial r_i}
		{\left(\displaystyle\sum_{i=1}^m\left(\frac{r_i}{r_{\max}}\right)^{p}\right)^{1 - \frac{1}{p}}}
		\quad = \quad  
		\frac{\displaystyle\sum_{i=1}^m -\left(\frac{r_i}{r_{\max}}\right)^{p-1}
			\left(\bm{\phi}_i^{\top}\left(\nabla \mathbf{G} + r_i \nabla \mathbf{K}\right) \bm{\phi}_i\right)}
		{\left(\displaystyle\sum_{i=1}^m\left(\frac{r_i}{r_{\max}}\right)^p\right)^{1 - \frac{1}{p}}}
	\end{equation}
	Here $\partial r_i$ corresponds to Clarke's generalized gradient of $r_i$~\cite{rodrigues1995necessary} and $\bm{\phi_i}$ are K-orthogonal eigenvectors. These aggregation functions uses a specific combination of symmetric polynomials~\cite{gravesen2011sensitivities}. That is why they are also smooth functions of the design variables. Further,  any eigenvector $\bm{\phi}$ belonging to the eigenvector space associated with repeated eigenvalues can be used to compute the gradient of aggregation functions. Indeed, aggregation functions are currently the most reliable method in addressing the non-differentiability of eigenvalues. However, a few challenges arise when applying these functions, especially in eigenvalue problems.
	
	\begin{enumerate}
		\item \textbf{Choice of aggregation parameter}: With a very small aggregation parameter, the corresponding aggregation function approximates $r_{\max}$ poorly. On the other hand, using a very large aggregation parameter makes the algorithm prone to convergence issues, often leading to inferior designs or getting stuck in undesirable local minima beyond a certain threshold. Additionally, the number of iterations required to achieve convergence will increase significantly under such conditions.
		\item \textbf{Choice of number of eigenvalues}: The final solution obtained through the aggregation function depends on the number of eigenvalues taken for the aggregation. When a very small number of eigenvalues is used, the gradients computed at multimodal points are incorrect; therefore, the final solution cannot be considered the optimized one. When a very large number of eigenvalues is used, computational time increases.
		\item \textbf{Modality at the optimum}: As the aggregation function approximates max($r_i$), the solution it produces may not be very close to the true solution. When the actual optimal solution is multimodal, the relative differences between the eigenvalues obtained using the aggregation method are very large, making it difficult to predict how many eigenvalues are coalescing. Without knowing the modality at the optimal point, it is not possible to find all the eigenvectors. Hence, we can't know all the possible buckled mode shapes.
	\end{enumerate}
	These issues are very well recognized in the literature. The effects of finite aggregation parameter on the optimal solution and its remedy have been discussed by Kennedy and Hicken~\cite{kennedy2015improved}. However, to the best of authors' knowledge, the second and third challenges have not yet been examined in depth. Addressing these two issues in the following sections and providing viable solutions is the primary goal of this paper.

\section{Choice of the number of eigenvalues}\label{Sec3:ChoiceEigen}

    As the modality at the optimum point is typically unknown, a relatively large number of eigenvalues--generally ranging from 10 to 30 are included in the aggregation function~\cite{torii2017structural,ferrari2019revisiting,ferrari2021topology,chin2016large}. With this approach, the computational cost increases as we compute a large number of eigenvalues and their corresponding eigenvectors at each iteration. Additionally, the optimal solution may have some accuracy issues, which we demonstrate in this section. Further, for a given problem,  even if the first eigenvalue is unimodal at a particular iteration, the approach still calculates the preselected number (a large number) of eigenvalues and eigenvectors. 

    First, we demonstrate, with an example, that choosing a fixed number of eigenvalues to aggregate in the optimization routine may be mathematically inconsistent, leading to incorrect gradient calculations. Thereafter, a novel approach to tackle this issue is proposed. Numerical experiments using the new approach are presented subsequently. 

\subsection{On number of eigenvalues to be aggregated}\label{sec:Noofeigentoagg}
     Let matrix $\mathbf{A}$ be defined as
	\[
	\mathbf{A} =
	\begin{bmatrix}
		0.99 & 0 & 0 & 0 \\
		0 & 0.99 & 0 & 0 \\
		0 & 0 & 1+x & y \\
		0 & 0 & y & 1-x
	\end{bmatrix}
	\]
One determines eigen spectrum of $\mathbf{A}$,
 $\bm{\Lambda} = \left\{\lambda_i\right\}|_{i =1,\,2,\,3,\,4} = \{0.99,0.99,  1 -\sqrt{x^2 + y^2}, 1 + \sqrt{x^2 + y^2}\}$. Here, $\lambda_i$ can take both positive and negative values; as per $x$ and $y$; thus, we set $r_i = -\lambda_i$. However, for a structural problem, one can always use $r_i = 1/\lambda_{i}$, as buckling load factors are always positive.
	
	At $(x,y) = (0,0),\quad \lambda_1 = \lambda_2 = 0.99 \quad \text{and} \quad \lambda_3 = \lambda_4 = 1$ which implies 
	$r_1 = -0.99, \hspace{0.1cm} r_2 = -0.99,\hspace{0.1cm} r_3 = -1,\hspace{0.1cm}r_4 = -1$. The eigenvector matrix $\bm{\Phi} = \left[\bm{\phi}_1,\,\bm{\phi}_2,\,\bm{\phi}_3,\,\bm{\phi}_4\right]= \mathbf{I}_4$, where $\mathbf{I}_4$ is an identity matrix of size $4\times 4$. The first eigenvalue is bimodal, while the third and fourth eigenvalues are closer to it at (0,\,0). Next, we present two cases illustrating the choice of the number of eigenvalues $m$ to be aggregated for evaluating the gradient of $C_\text{KS}$ (Eq.~\ref{Eq:KS_function}).
    

\subsubsection{CASE I}\label{sec:CASEI_repeateigen}
    In this case, $m=3$ is chosen to determine the gradient of $C_\text{KS}$, i.e., we aggregate the first three eigenvalues.  With $m=3$, Eq.~\ref{Eq:KS_der} yields as:
	\begin{equation}\label{Eq:Example1}
		\nabla C_{\mathrm{KS}}(r_i, \rho)=
\frac{\displaystyle \sum_{i=1}^{m=3}
\exp\!\left( \rho (r_i - r_{\max}) \right)\, \partial r_i}
{\displaystyle \sum_{i=1}^{m=3}
\exp\!\left( \rho (r_i - r_{\max}) \right)}
=
\frac{\displaystyle \sum_{i=1}^{m=3}
\exp\!\left( \rho (r_i - r_{\max}) \right)
\left( -\, \bm{\phi}_i^{\top} \, (\nabla \mathbf{A}) \, \bm{\phi}_i \right)}
{\displaystyle \sum_{i=1}^{m=3}
\exp\!\left( \rho (r_i - r_{\max}) \right)}
	\end{equation}
		$\text{where} \hspace{0.2cm} \nabla\mathbf{A} = \left\{\frac{\partial \mathbf{A}}{\partial x}, \frac{\partial \mathbf{A}}{\partial y}\right\}$. Upon expanding the LHS of Eq.~\ref{Eq:Example1}, we get

\begin{equation}\label{Eq:Example1_1}
    \begin{split}
 \nabla C_{\mathrm{KS}}\bigl(r_i(0,0), \rho\bigr)
&=
\frac{\exp\!\bigl(\rho(-0.99+0.99)\bigr)(0)+\exp\!\bigl(\rho(-0.99+0.99)\bigr)(0)+\exp\!\bigl(\rho(-1+0.99)\bigr)
\left(-\, \bm{\phi}_3^{\top} (\nabla \mathbf{A}) \bm{\phi}_3 \right)}{\exp\!\bigl(\rho(-0.99+0.99)\bigr)+\exp\!\bigl(\rho(-0.99+0.99)\bigr)+\exp\!\bigl(\rho(-1+0.99)\bigr)}\\
&=
\frac{
-\, \exp(-0.01\,\rho)\,
\bm{\phi}_3^{\top} (\nabla \mathbf{A}) \bm{\phi}_3
}{
2 + \exp(-0.01\,\rho)
}
    \end{split}
\end{equation}

    $\lambda_3$ and $\lambda_4$ are also repeating eigenvalues.
	 Post-multiplying $[\bm{\phi}_3,\bm{\phi_4}]$ with an arbitrary orthogonal matrix~G gives a new orthogonal basis for the eigenspace corresponding to the repeated eigenvalues. Take $G = \begin{bmatrix}
		\cos\alpha & \sin\alpha\\
		-\sin\alpha & \cos\alpha
	\end{bmatrix}$, then the determined new eigenvectors  are $[\bm{\phi}_3,\bm{\phi_4}] = [(0,\,0,\,\cos\alpha,\,-\sin\alpha)^\top, (0,\,0,\,\sin\alpha,\,\cos\alpha)^\top]$, where $\alpha \in [0,2\pi)$. Per~\cite{torii2017structural}, as any orthogonal basis can be used in computing the gradient of the aggregation function, we use the new form of $\bm{\phi}_3$ eigenvector and find
	$\bm{\phi}_3^{\top}\left(\nabla \mathbf{A}\right) \bm{\phi}_3 = \begin{bmatrix}
		\bm{\phi}_3^{\top}\left(\frac{\partial \mathbf{A}}{\partial x}\right) \bm{\phi}_3 \\ \bm{\phi}_3^{\top}\left(\frac{\partial \mathbf{A}}{\partial y}\right) \bm{\phi}_3
	\end{bmatrix}
	= \begin{bmatrix}
		\cos (2\alpha) \\
		-\sin (2\alpha)
	\end{bmatrix}$, which transpire Eq.~\ref{Eq:Example1_1} to
\begin{equation}\label{Eq:Del_KS_1}
        \nabla C_{\mathrm{KS}}\bigl(r_i(0,0), \rho\bigr)=
\frac{-\,\exp(-0.01\,\rho)}{2 + \exp(-0.01\,\rho)}
	\begin{bmatrix}
		\cos (2\alpha) \\
		-\sin (2\alpha)
	\end{bmatrix}  
\end{equation}
Eq.~\ref{Eq:Del_KS_1} reveals that $\nabla C_{\mathrm{KS}}\bigl(r_i, \rho\bigr)$ depends on the choice of the eigenvector basis of the repeated eigenvalue, which should not be the case. Therefore, the computed gradient is incorrect for the selected $m=3$.


\subsubsection{CASE II}
  In this case, we take $m=4$  and repeat the steps mentioned in Sec.~\ref{sec:CASEI_repeateigen} (CASE I) for Eq.~\ref{Eq:Example1} and Eq.~\ref{Eq:Example1_1}, we get
	\begin{equation}\label{Eq:caseII_cal}
    \begin{split}
		\nabla C_{\mathrm{KS}}\bigl(r_i(0,0), \rho\bigr)
&=
\frac{
-\,\exp(-0.01\,\rho)\,
\left[
\bm{\phi}_3^{\top} (\nabla \mathbf{A}) \bm{\phi}_3
+
\bm{\phi}_4^{\top} (\nabla \mathbf{A}) \bm{\phi}_4
\right]
}{
2\left(1 + \exp(-0.01\,\rho)\right)
}\\
&=
\frac{-\,\exp(-0.01\,\rho)}{2\left(1 + \exp(-0.01\,\rho)\right)} \left(
		\begin{bmatrix}
			\cos(2\alpha) \\
			-\sin(2\alpha)
		\end{bmatrix} + 
		\begin{bmatrix}
			-\cos(2\alpha) \\
			\sin(2\alpha)
		\end{bmatrix} \right)= \begin{bmatrix}
			0 \\
			0
		\end{bmatrix}
\end{split}
	\end{equation}	
Equation~\ref{Eq:caseII_cal} indicates that the gradient of the aggregation function is independent of $\alpha$ or the eigenvector basis. Therefore, $m$ should not be a random choice for the correct calculation of the derivative of aggregation functions. 

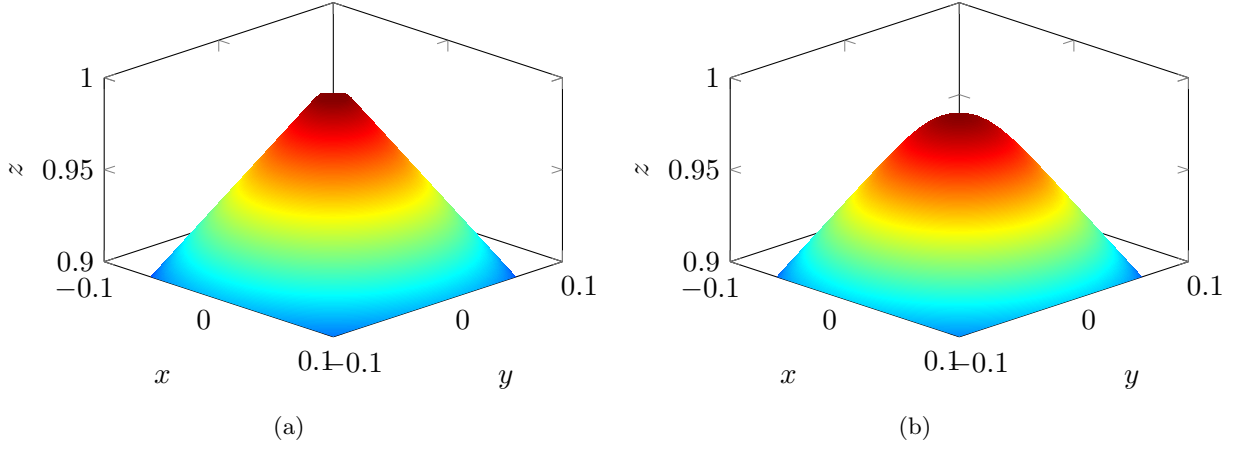
\begin{figure}
\centering

\begin{subfigure}[t]{0.48\textwidth}
\centering

\begin{tikzpicture}
\begin{axis}[
    view={45}{30},
    xlabel={$x$},
    ylabel={$y$},
    zlabel={$z$},
    xmin=-0.1,xmax=0.1,
    ymin=-0.1,ymax=0.1,
    zmin=0.9,zmax=1,
    width=\linewidth,
    height=6cm,
    colormap/jet,
]

\addplot3[
    surf,
    shader=interp,
]
table[
    x index=0,
    y index=1,
    z index=2,
] {cone1.txt};

\end{axis}
\end{tikzpicture}

\caption{}
\label{fig:cone 1}

\end{subfigure}
\hfill
\begin{subfigure}[t]{0.48\textwidth}
\centering

\begin{tikzpicture}
\begin{axis}[
    view={45}{30},
    xlabel={$x$},
    ylabel={$y$},
    zlabel={$z$},
    xmin=-0.1,xmax=0.1,
    ymin=-0.1,ymax=0.1,
    zmin=0.9,zmax=1,
    width=\linewidth,
    height=6cm,
    colormap/jet,
]

\addplot3[
    surf,
    shader=interp,
]
table[
    x index=0,
    y index=1,
    z index=2,
] {cone2.txt};

\end{axis}
\end{tikzpicture}

\caption{}
\label{fig:cone 2}

\end{subfigure}

\caption{(a) The graph of $\min\left\{0.99,0.99, 1 - \sqrt{x^2 + y^2}, 1 + \sqrt{x^2 + y^2}\right\}$ (b) The graph of $C_\text{KS}(r_{i},\rho)$ where $r_i = -\lambda_i$ $i = 1,2,3,4$ and $\rho = 100$.}
\label{fig:aggregationexample}
\end{figure}

Next, the plots for  $\min\left\{0.99,0.99, 1 - \sqrt{x^2 + y^2}, 1 + \sqrt{x^2 + y^2}\right\}$ and $C_\text{KS}(r_{i},\rho)$ where $r_i = -\lambda_i$ $i = 1,2,3,4$ and $\rho = 100$ are depicted in Fig.~\ref{fig:cone 1} and Fig.~\ref{fig:cone 2}, respectively. (0,\,0) indicate an optimal point (Fig.~\ref{fig:aggregationexample}).  
	
In the example presented above, the modality of the first eigenvalue is two. We might assume that aggregating more than two eigenvalues ($m >= 2$) will solve the problem. However, the third and fourth eigenvalues also coalesce at the optimum point wherein $\lambda_1 = \lambda_2$ and $\lambda_3 = \lambda_4$, but $\lambda_1 \neq \lambda_3$. Even though the third eigenvalue does not coalesce with the first, calculating the gradient using three eigenvalues ($m = 3$) gives an incorrect result (Sec.~\ref{sec:CASEI_repeateigen}). As shown in Eq.~\ref{Eq:Del_KS_1}, this gradient error becomes especially large when the gap $\lambda_1 - \lambda_3$ is very small. The only way to get correct gradients is to choose either $m = 2$ or $m = 4$. Additionally, remark that the gradient for $m = 3$ is incorrect only at the point $(x,y) = (0,0)$, because this is the only location where $\lambda_3 = \lambda_4$ strictly holds.

This example demonstrates that choosing an arbitrary number of eigenvalues leads to incorrect gradients. Specifically, for a problem with eigenvalue multiplicity $q$, choosing $m > q$ produces incorrect gradients whenever the $m$-th eigenvalue merges with higher-order ones. Therefore, the number of aggregated eigenvalues must be selected dynamically at each iteration based on their relative separations.


\subsection{$\varepsilon$-ln threshold difference (ELTD) approach}
   
	If one selects $m$ eigenvalues to aggregate, then at any iteration, there could be a possibility that $m$-th and the next eigenvalues might coalesce. In such a case, whatever derivative one calculates will be incorrect. This is because when a coalescing eigenvalue is not considered, one can't use an arbitrary eigenvector basis to compute sensitivities as illustrated in Sec.~\ref{sec:Noofeigentoagg}. If one wants to choose any arbitrary eigenvector while computing the sensitivity of the aggregation function, then all the repeated eigenvalues must be considered in the aggregation function. We demonstrate in Sec.~\ref{sec:Noofeigentoagg} that even if the problem is bimodal and the number of eigenvalues aggregated is 3, the sensitivities one computes are incorrect. Additionally, in a TO setting, $q$ may vary across iterations, which makes problems more pronounced. Therefore, we need a novel scheme to determine $m$ for the given problem in each iteration of the optimization process.

    Assume $r_{1} \ge r_{2}..... \ge r_{m}$,, i.e., $r_{\max} = r_1$, then  Eq.~\ref{Eq:KS_function} can be written as
    	\begin{equation}\label{Eq:KS_function_r1=rmax}
C_{\mathrm{KS}}(r_i, \rho)  = r_1 + \frac{1}{\rho} \ln \left( 1 + \sum_{i=2}^{m} \exp\!\left( \rho (r_i - r_1) \right) \right)
	\end{equation}
    and Eq.~\ref{Eq:KS_der} yields as
    \begin{equation}\label{Eq:der_KS_r1=rmax}
    \begin{split}
        \nabla C_{\mathrm{KS}}(r_i, \rho) =& \frac{\partial r_1 + \sum_{i=2}^{m} \exp\!\left( \rho (r_i - r_1) \right)\, \partial r_i}{1 + \sum_{i=2}^{m} \exp\!\left( \rho (r_i - r_1) \right)}\\
        =& \frac{\partial r_1 + \exp\!\left( \rho (r_2 - r_1) \right)\, \partial r_2 + \exp\!\left( \rho (r_3 - r_1) \right)\, \partial r_3 + \cdots + \exp\!\left( \rho (r_m - r_1) \right)\, \partial r_m}{1 + \exp\!\left( \rho (r_2 - r_1) \right) + \exp\!\left( \rho (r_3 - r_1) \right) + \cdots + \exp\!\left( \rho (r_m - r_1) \right)}\\
        =&\frac{\partial r_1 + w_2\, \partial r_2 + w_3\, \partial r_3 + \cdots + w_m\, \partial r_m}{1 + w_2 + w_3 + \cdots + w_m},
    \end{split}
    \end{equation}
where $w_m$ is referred to as the relative KS weight of mode $m$.

To the best of authors' knowledge, the above-mentioned issue (to determine how many eigenvalues to be aggregated in every optimization iteration) is neglected because, when a large number of eigenvalues is aggregated, the difference between the first, $r_1$, and last $r_m$, eigenvalues is usually large. As a result, the coefficient of sensitivity or KS weight of the last eigenvalue (the exponential) is relatively very small (Eq.~\ref{Eq:der_KS_r1=rmax}). Any errors in the sensitivities of the last eigenvalue don't affect the overall gradient of the aggregation function (Eq.~\ref{Eq:der_KS_r1=rmax}). That is, when the separation between the first and last eigenvalues, as well as the aggregation parameter $\rho$, is large, this issue numerically doesn't affect the final sensitivities (Eq.~\ref{Eq:der_KS_r1=rmax}); thus, the optimized solution since $w_m\to 0$. That is why one typically chooses a large $m$ to circumvent this issue. However, the questions we ask are: \textit{how large should $m$ be, and how should it be determined to evaluate correct sensitivity}? Additionally, the required computational time is directly proportional to $m$. Therefore, a method is required to systematically guide us in selecting $m$ for the given problem. Herein, we present the $\varepsilon$-ln threshold difference approach, which selects $m$ based on the modality at each iteration, without causing abrupt changes to the objective or constraint functions.

     Per Eq.~\ref{Eq:der_KS_r1=rmax}, $\exp\!\left( \rho (r_i - r_1) \right)=w_i<1$, as $ (r_i-r_1)<0$  and $\rho>0$, i.e.,  contributions of $w_i$ in the summation are guided by the terms,  $(r_i-r_1)$  and $\rho$. For a very high, $(r_i-r_1)$ or $\rho$ or $\rho(r_i-r_1)$;  $w_i<<1$, indicating its contribution in the derivative terms can be neglected.  If any $w_i<\varepsilon$; where $\varepsilon<<1$, that is
     \begin{equation} \label{Eq:epsilon-ln}
\begin{aligned}
w_i &< \varepsilon, \implies
(r_1 - r_i) &> -\frac{\ln(\varepsilon)}{\rho}
\end{aligned}
\end{equation}
then  one can negelect it in Eq.~\ref{Eq:der_KS_r1=rmax}. The expression in Eq.~\ref{Eq:epsilon-ln} is termed the $\varepsilon$-ln threshold difference (ELTD) scheme, as it depends upon the difference $(r_i - r_1)$ and $\ln(\varepsilon)$. The numerical experiments presented in Appendix~\ref{app:app1} confirm that $\epsilon=\num{1e-9}$ works well, as it prevents sudden jumps in the objective or constraint functions and captures eigenvalue repetition smoothly. However, one can take $\varepsilon$ even a little higher (Appendix~\ref{app:app1}). 

\begin{figure}

    \begin{subfigure}[t]{0.45\textwidth}
    
        \begin{tikzpicture}
            \begin{axis}[
                xlabel={$x$},
                ylabel={$\lambda$},
                xmin=-2,xmax=2,
                ymin=0,ymax=4,
                width=\linewidth,
                legend style={
                    at={(0.5,0.2)},
                    anchor=south west,
                    font=\small
                }
            ]

            \addplot[red,solid,line width=1.5pt]
            table[x index=0,y index=1]
            {multimodal_limits_1_and_2_a1.txt};

            \addlegendentry{
            $f_0$}

            \addplot[black,dashed,line width=1.5pt]
            table[x index=0,y index=1]
            {multimodal_limits_1_a2.txt};

            \addlegendentry{$C_\text{KS}^1$}

            \addplot[blue,dashdotted,line width=1.5pt]
            table[x index=0,y index=1]
            {multimodal_limits_1_a3.txt};
             \addlegendentry{$x_1$}
            \addplot[green,dashdotted,line width=1.5pt]
            table[x index=0,y index=1]
            {multimodal_limits_1_a4.txt};
             \addlegendentry{$x_2$}

            \end{axis}
        \end{tikzpicture}
        \caption{}
    \end{subfigure}
    \begin{subfigure}[t]{0.45\textwidth}
    \centering
        \begin{tikzpicture}
            \begin{axis}[
                xlabel={$x$},
                ylabel={$\lambda$},
                xmin=-2,xmax=2,
                ymin=0,ymax=4,
                width=\linewidth,
                legend style={
                    at={(0.5,0.2)},
                    anchor=south west,
                    font=\small
                }
            ]

            \addplot[red,solid,line width=1.5pt]
            table[x index=0,y index=1]
            {multimodal_limits_1_and_2_a1.txt};

            \addlegendentry{$f_0$}

            \addplot[black,dashed,line width=1.5pt]
            table[x index=0,y index=1]
            {multimodal_limits_2_a2.txt};
            \addlegendentry{$C_\text{KS}^2$}

            \addplot[blue,dashdotted,line width=1.5pt]
            table[x index=0,y index=1]
            {multimodal_limits_2_a3.txt};
             \addlegendentry{$x_3$}

            \addplot[green,dashdotted,line width=1.5pt]
            table[x index=0,y index=1]
            {multimodal_limits_2_a4.txt};
             \addlegendentry{$x_4$}
 
            \end{axis}
        \end{tikzpicture}
        \caption{}
    \end{subfigure}

    \caption{The graph showing the minimum of two functions \{$\lambda_1$,$\lambda_2$\} and their KS aggregation function approximation for two aggregation parameters (a) $\rho = 5$ and (b) $\rho = 100$.  
            $f_0=\min(\lambda_1 = 2x+3,\,\lambda_2 = -x^2 + 4)$, $C_\text{KS}^1= C_\text{KS}\!\left(\frac{1}{\lambda_1},\frac{1}{\lambda_2},\rho=5\right)^{-1}$, and $C_\text{KS}^2= C_\text{KS}\!\left(\frac{1}{\lambda_1},\frac{1}{\lambda_2},\rho=100\right)^{-1}$}
        \label{fig:multimodal limits}
\end{figure}
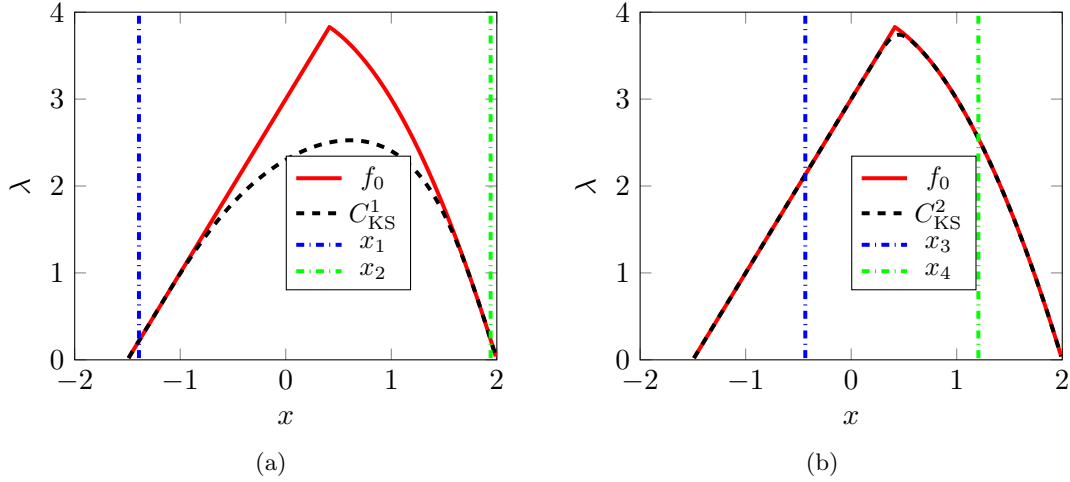

	The physical intuition behind $\varepsilon$-ln threshold difference scheme for choosing the number of eigenvalues is demonstrated using
	Fig.~\ref{fig:multimodal limits}. Consider an eigenvalue spectrum whose eigenvalues are  $\lambda_1 = 2x+3 \text{ and } \lambda_2=-x^2 + 4$. $\lambda_1$ and $\lambda_2$ intersect at $x_i = 0.41421$. At this point, the function $\lambda_{\min}= \min(\lambda_1,\lambda_2)$ is not differentiable as $\lambda_{\min}$ changes from $\lambda_1$ to $\lambda_2$. Now, we approximate the function $\lambda_{\min}$ using KS-aggregation. Far away from the point $x_i$, the aggregation function behaves like either $\lambda_1$ or $\lambda_2$. That is for $x<<x_i$, $C_{KS}\left(\frac{1}{\lambda_1},\frac{1}{\lambda_2},\rho\right)^{-1} \approx \lambda_1$ and $x>>x_i$, $C_{KS}\left(\frac{1}{\lambda_1},\frac{1}{\lambda_2},\rho\right)^{-1} \approx \lambda_2$. Thus, far away from the bimodal point, we can just use one eigenvalue; that is, no aggregation is needed. Near the vicinity of $x_i$, to capture the sharp change in gradient, we need to use both eigenvalues for aggregation. The region where both eigenvalues are necessary is found using the $\varepsilon$-ln threshold difference scheme for two cases of $\rho$ as illustrated in Fig.~\ref{fig:multimodal limits}. We take $\varepsilon = 10^{-9}$ to mark the regions of aggregation, indicated via the green and blue dash-dotted lines in Fig.~\ref{fig:multimodal limits}, as per Eq.~\ref{Eq:epsilon-ln}. These green and blue dash-dotted lines also highlight the fact that $\varepsilon= 10^{-9}$ is on the conservative side.
    
If the modality of the eigenvalue is say $q$, ideally, we only need to aggregate $q$ eigenvalues for approximating $r_{\max}$. The more eigenvalues one considers in the aggregation, the more the error will be. Using the presented ELTD scheme in conjunction with the KS aggregation function, one can approximate $\max(r_i)$ accurately by selecting dynamically only those eigenvalues that are very close to the first eigenvalue. Using this approach, one calculates only a few eigenvalues per iteration, thereby saving significant computation time without compromising the optimized solution. Next, we describe how to select the adaptive value of $m$ in each iteration for the proposed ELTD approach.

	\begin{figure}[htbp]
		\centering
		\begin{tikzpicture}[node distance=2.2cm]
			\node (start) [process] {
				Calculate $n_{\text{agg}}^{(i)} = n_{\text{agg}}^{(i-1)} + 1$ eigenvalues
			};
			
			\node (findk) [process, below of=start] {
				Find index $k$ such that \\
				$(r_1 - r_k) > -\ln(\varepsilon)/\rho$
			};
			\node (pass) [process, right=3.5cm of findk] {Then pass $(k-1)$\\ eigenvalues into\\ the aggregation function};

			\node (update) [process, below of=findk] {
				Calculate $n_{\text{agg}}^{(i)} = n_{\text{agg}}^{(i)} + 2$ eigenvalues \\
				(including previously found eigenvalues)
			};
			
			\draw [arrow] (start) -- (findk);
			
			\draw [arrow] (findk) -- node[right]{(If no $k$ exists)} (update);
			
			\draw [arrow] (findk.east) -- node[above]{(If such $k$ exists)} (pass.west);
			
			\draw [arrow] (update.west) -| ++(-2.2,2.2) |- (findk.west);
		\end{tikzpicture}
		\caption{Flowchart explaining how eigenvalues are passed onto the aggregation function in each iteration for the ELTD approach. $n_{\text{agg}}^{(i)}$ indicates the number of eigenvlaues aggregated in the $i$-th iteration.}
		\label{fig:flowchart}
	\end{figure}
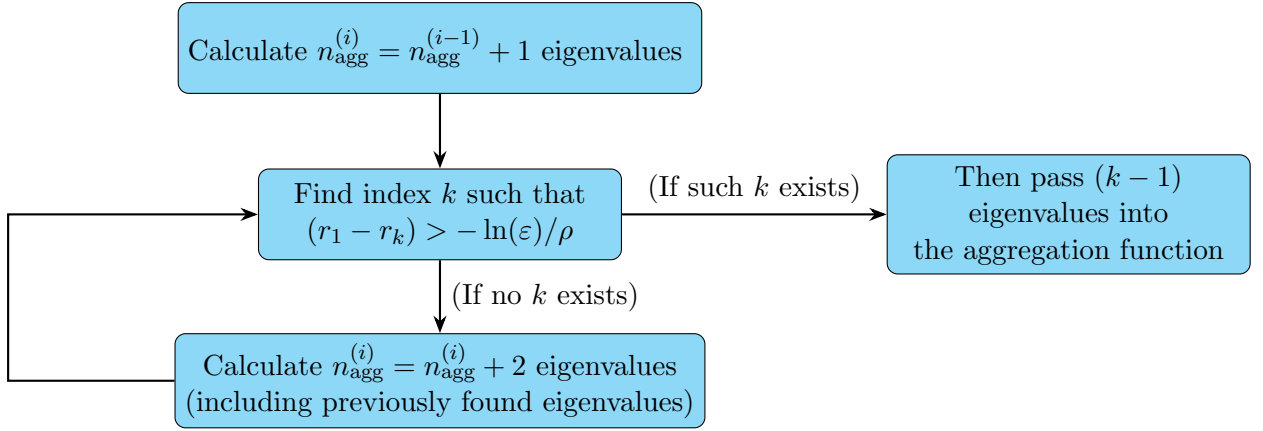

 In the $i$-th iteration, we begin by determining $n_\text{agg}^{(i-1)} + 1$ eigenvalues, where $n_\text{agg}^{(i-1)}$ is the number of eigenvalues passed to aggregation function in ${(i-1)}$-th iteration. If the $k$-th eigenvalue satisfies Eq.~\ref{Eq:epsilon-ln}, then the $k-1$ eigenvalues are sent to the aggregation function. If any of the calculated eigenvalues doesn't satisfy  Eq.~\ref{Eq:epsilon-ln}, then we calculate $n_\text{agg}^{(i)} = n_\text{agg}^{(i)} + 2$ eigenvalues and see whether the newly found 2 eigenvalues satisfy Eq.~\ref{Eq:epsilon-ln}. The process continues until we find a $k$-th eigenvalue. We use the \texttt{eigs} MATLAB function to determine eigenvalues. Note that when the first set of eigenvalues fails to satisfy the criteria (Eq.~\ref{Eq:epsilon-ln}), we calculate the entire set of $n_{agg}^{(i)} + 2$ eigenvalues, but not only two extra eigenvalues.

	The \texttt{eigs} function in MATLAB  uses the iterative Krylov-Schur algorithm (Arnoldi process)~\cite{stewart2002krylov}. It builds a Krylov subspace of dimension $\max(2p,20)$ by default, where $p$ is the desired number of eigenvalues that need to be calculated. The algorithm runs for up to 300 iterations, or until the prescribed tolerance is achieved, and does not compute eigenvectors sequentially. It searches for all eigenvectors of the constructed subspace simultaneously, computing them on the fly. In our case, whenever we cannot find the index $k$, we need to find two additional eigenvalues using the routine described above; see also the provided flowchart in Fig.~\ref{fig:flowchart}. To find the extra two eigenvalues, we need to $K-$orthonormalize the previous eigenvectors and give a deflated vector to the matrix function handle in the \texttt{eigs} function. This leads to very slow convergence. That is why it is more time efficient to find all the $n_\text{agg}^{(i)} + 2$ eigenvalues once again. Additionally, as the eigenvalues approach each other, the algorithm takes longer to converge. In such a case, it is often more time-efficient to call all the very close eigenvalues. This is because when a large number of eigenvalues are requested, the number of restarts required will be less, and all the eigenvalues will be converged in one go in the Arnoldi (implicitly restarted) method~\cite{lehoucq1998arpack}. In this paper, this phenomenon is observed for Example 2 after 400 iterations (See sec.~\ref{Sec:Example2ELTD}, depicted in Fig.~\ref{Fig:P2_Time_variation}). At that location, the time required by the conventional method to compute 12 eigenvalues is less than that of the ELTD approach, which computes  6 eigenvalues.

 \subsection{Numerical examples and discussion}
    To demonstrate the efficacy, robustness, and success of the $\varepsilon$-ln threshold difference scheme, the axially compressed column~\cite{ferrari2021topology,gao2015topology,bian2017large,gao2017adaptive, pedersen2018buckling}, wall reinforcement~\cite{ferrari2021topology}, and shear-loaded hinged plate ~\cite{ferrari2019revisiting,ferrari2020towards} problems are investigated in this section.

\subsubsection{Axially compressed column} \label{Sec:Example1ELTD}
\begin{figure}[h!]
		\centering
		\begin{subfigure}[c]{0.45\textwidth}
			\centering
			\includegraphics[width=\textwidth]{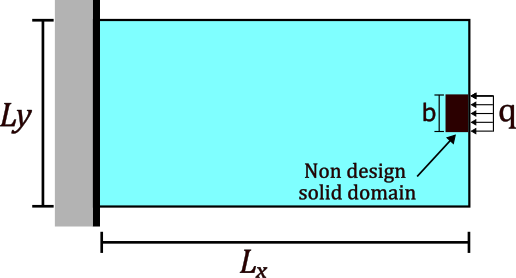} 
			\caption{}
			\label{Fig:P1_Schematic}
		\end{subfigure}
		\hfill
		\begin{subfigure}[c]{0.45\textwidth}
			\centering
			\includegraphics[width=\textwidth]{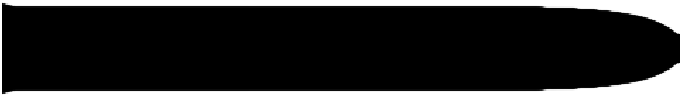} 
			\caption{}
			\label{Fig:P1_no_buck_design}
		\end{subfigure}
		\caption{(a) Design domain for axially compressed column and (b) Minimum volume design}\label{Fig:P1_Schematic_and_no_buck_design}
	\end{figure}

An axially compressed 2D column with $L_x = 2$ and $L_y = 1$ is depicted in Fig.~\ref{Fig:P1_Schematic}.  The left edge is fixed, while a surface load intensity $q=\num{1e-3}$ is acting near the center of the right edge over a width of $b = \frac{L_y}{15}$. The domain is parameterized using $480 \times 240$ bilinear rectangular FEs. A non-design solid domain with $\frac{L_x}{48}\times\frac{L_y}{12}$ FEs is considered at the applied force location (Fig.~\ref{Fig:P1_Schematic}). The optimization problem solved is framed as:
    \begin{equation}\label{Eq:Axially_Opt_prolem}
        \begin{aligned}
		&\min_{\bm{x}} \quad f({\bm{x}})/{f_0} \\
		&\text{s.t.} \quad g_\text{C}({\bm{x}}) = C({\bm{x}})/\bar{C} - 1 \leq 0
	\end{aligned}
    \end{equation}
	${\bm{x}}$, $f({\bm{x}})$, and  ${f_0}$ represent the design vector, volume of the structure, and total volume of the initial structure, respectively. $C(\bm{x})$ and $\bar{C}$ indicate the structure's compliance and the permitted compliance, respectively. The minimum volume design subjected to compliance constraint of $\bar{C} = 2.5C_{\text{loop} = 1}$ is obtained using \texttt{TopBuck250} code \cite{ferrari2021topology}, where $C_{\text{loop} = 1}$ indicates the compliance value at the first optimization iteration (loop =1).  The final optimized design obtained from Eq.~\ref{Eq:Axially_Opt_prolem} is shown in Fig.~\ref{Fig:P1_no_buck_design}. It has a volume fraction of 0.2408, $g_C = -4.2253\times10^{-7}$, $C^* = \num{8.8407e-06}$ (compliance at the optimum point) and a buckling load of 0.7823.
	
In the second step, we maximize its critical buckling load subjected to compliance and volume constraints. For that, the optimization problem is formulated as:
    
\begin{equation}\label{Eq:BucklingProblem_1}
\begin{aligned}
\min_{\bm{x}} \quad & C_{\mathrm{KS}}[r_i](\bm{x}) \\
\text{s.t.} \quad & g_\text{C}({\bm{x}}) = C({\bm{x}})/\bar{C} - 1 \leq 0\\
\quad \,\, &g_V(\bm{x}) = \frac{f(\bm{x})}{\bar{f}} - 1 \leq 0
\end{aligned}
\end{equation}
 The final design depicted in Fig.~\ref{Fig:P1_no_buck_design} is given as an initial guess to the buckling problem (Eq.~\ref{Eq:BucklingProblem_1}), subjected to the volume constraint of $\bar{f} = 0.25\times\text{total volume of the structure}$ and compliance constraint of $\bar{C} = 1.05C^*$.  Per~\cite{ferrari2021topology}, both constraints become active at the optimum point for this problem. We use the Method of Moving Asymptotes (MMA)~\cite{svanberg1987method} algorithm herein to update the design variables.
	\begin{figure}
		\centering
		\begin{subfigure}[b]{0.45\textwidth}
			\centering
			\includegraphics[width=\textwidth]{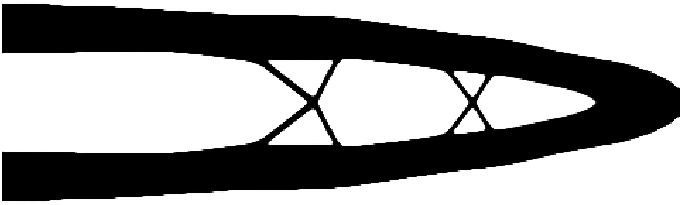} 
			\caption{}
			\label{Fig:P1_conven_method_design}
		\end{subfigure}
		\hfill
		\begin{subfigure}[b]{0.45\textwidth}
			\centering
			\includegraphics[width=\textwidth]{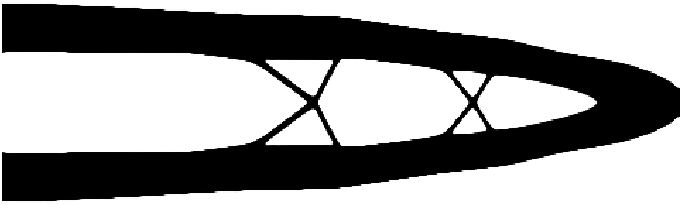} 
			\caption{}
			\label{Fig:P1_ELTD_method_design}
		\end{subfigure}
		\caption{(a) Conventional method: Maximum buckling load design obtained by aggregating 12 eigenvalues in all iterations and (b) ELTD approach: Maximum buckling load design obtained by dynamically aggregating only a few eigenvalues} \label{Fig:P1_conven_and_ELTD_buck_design}
	\end{figure} 

\begin{table}[H]
\centering
\caption{Comparison of the conventional and ELTD approaches for the compressed column problem.}
\label{Tab:P1_final_results}
\renewcommand{\arraystretch}{1.15}
\setlength{\tabcolsep}{8pt}

\begin{tabular}{l |c |c}
\hline\hline
Physical quantity &
Conventional approach ($m=12$) &
ELTD approach ($\varepsilon=10^{-9}$)
\\
\hline

Minimum BLF &
4.9650 &
4.9894
\\

$g_C$ &
$-1.45\times10^{-3}$ &
$-1.90\times10^{-5}$
\\

$g_V$ &
$-5.16\times10^{-4}$ &
$-1.40\times10^{-4}$
\\

Second BLF &
5.2422 &
5.2540
\\

Computational time (s) &
9130.36 &
4634.83
\\

\hline\hline
\end{tabular}
\end{table}


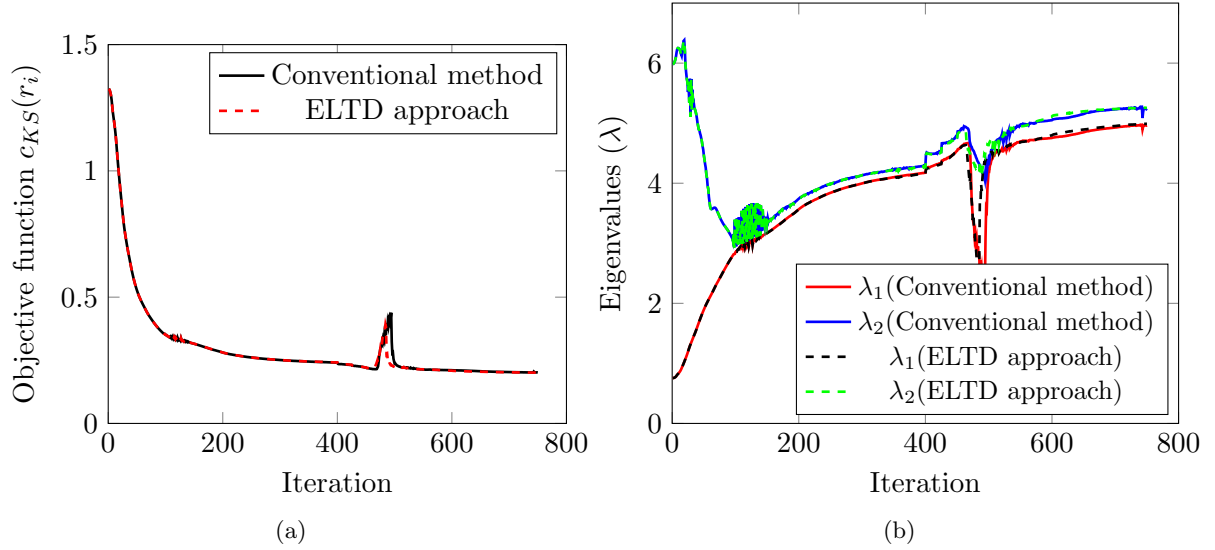
\begin{figure}[h!]
		\begin{subfigure}[t]{0.48\textwidth}
        \centering
			\begin{tikzpicture}
				\begin{axis}[
					xlabel={Iteration},
					ylabel={Objective function $c_{KS}(r_i)$},
					xmin=0,xmax=800,
					ymin=0, ymax=1.5,
                    width=\linewidth,
					]    
					\addplot[solid,{black}, line width=1pt] table[x index=0, y index=1] {P1_Obj_val_variation_conven_method.txt}; \addlegendentry{Conventional method}
					\addplot[dashed,{red}, line width=1pt, mark = none] table[x index=0, y index=1] {P1_Obj_val_variation_ELTD_method.txt};
					\addlegendentry{ELTD approach}
				\end{axis}
			\end{tikzpicture}
			\caption{}
        \label{Fig:P1_Obj_val_variation}
		\end{subfigure}
		\begin{subfigure}[t]{0.52\textwidth}
        \centering
			\begin{tikzpicture}
				\begin{axis}[
					xlabel={Iteration},
					ylabel={Eigenvalues $(\lambda)$},
					xmin=0,xmax=800,
					ymin=0, ymax=7,
                    width=\linewidth,
                    legend style={ at={(0.98,0.02)},
                    anchor=south east,
                    font=\small}
					]    
					\addplot[solid,{red}, line width=1pt] table[x index=0, y index=1] {P1_Eig_val_1_variation_conven_method.txt}; \addlegendentry{$\lambda_1$(Conventional method)}
					\addplot[solid,{blue}, line width=1pt, mark = none] table[x index=0, y index=1] {P1_Eig_val_2_variation_conven_method.txt};
                     \addlegendentry{$\lambda_2$(Conventional method)}
                    \addplot[dashed,{black}, line width=1pt, mark = none] table[x index=0, y index=1] {P1_Eig_val_1_variation_ELTD_method.txt};
                    \addlegendentry{$\lambda_1$(ELTD approach)}
                    \addplot[dashed,{green}, line width=1pt, mark = none] table[x index=0, y index=1] {P1_Eig_val_2_variation_ELTD_method.txt};
                    \addlegendentry{$\lambda_2$(ELTD approach)}
                    
				\end{axis}
			\end{tikzpicture}
			\caption{}
        \label{Fig:P1_Eig_val_variation}
		\end{subfigure}
		\caption{Compressed Column problem: (a) Evolution of the objective function (b) Evolution of the first and second eigenvalues.}
        \label{Fig:P1_Obj_and_Eig_val_variation}
	\end{figure}

 The buckling problem (Eq.~\ref{Eq:BucklingProblem_1}) is solved using both the proposed ELTD approach and the conventional approach. For the former, $m$ is dynamically updated during the optimization process, wherein  $m=12$~\cite{ferrari2021topology}  is kept constant throughout the iterations for the latter. The optimized designs obtained using both approaches are depicted in Fig.~\ref{Fig:P1_conven_and_ELTD_buck_design}, which resemble each other. The conventional approach provides a minimum buckling load factor of  4.9650, with constraints' value $g_C = \num{-1.45e-3} \text{ and } g_V = \num{-5.16e-4}$, and the second load factor of 5.2422. This indicates that the optimized design is unimodal; however, 12 eigenvalues are aggregated per iteration. Using the ELTD approach with $\varepsilon = \num{1e-9}$, we obtain minimum buckling load factor of 4.9894 along with $g_C = \num{-1.90e-5} \text{ and } g_V = \num{-1.40e-4}.$ The second buckling load factor is 5.254. This problem is unimodal for almost all iterations, so aggregating many eigenvalues yields a poor approximation of $r_{max}$. This is supported by the fact that the optimal solution predicted by the ELTD solution has a larger $\lambda_1$ and satisfies the constraints more strictly than those of the conventional method. Table~\ref{Tab:P1_final_results} summarizes different evaluated parameters using both approaches, along with the required computational time. 

 The problem is solved on a PC with 12th Gen Intel(R) Core(TM) i7-12700K (3.60 GHz), 32GB RAM, and MATLAB R2024b. We observe that the ELTD approach takes 4634.83 sec, whereas the conventional method takes 9130.36 sec, resulting in approximately 49.24\% savings in computational time (Table~\ref{Tab:P1_final_results}). The computational time reported in the paper corresponds only to the time required to compute eigenvalues during the optimization routine. Additionally, we observe no fluctuations in the objective function (Fig.~\ref{Fig:P1_Obj_val_variation}) as the design modality changes during optimization. The objective function and eigenvalue convergence curves for both approaches are depicted in Fig.~\ref{Fig:P1_Obj_val_variation} and Fig.~\ref{Fig:P1_Eig_val_variation}, respectively. Both methods follow the same convergence paths, while the proposed approach saves significant computational time, as indicated above (Table~\ref{Tab:P1_final_results}). Next, we solve a wall reinforcement problem in which multiple eigenvalues coalesce at the optimal point.
	
\begin{figure}[h!]
		\centering
		\begin{subfigure}[c]{0.45\textwidth}
			\centering
			\includegraphics[height=8cm,width=\textwidth,keepaspectratio]{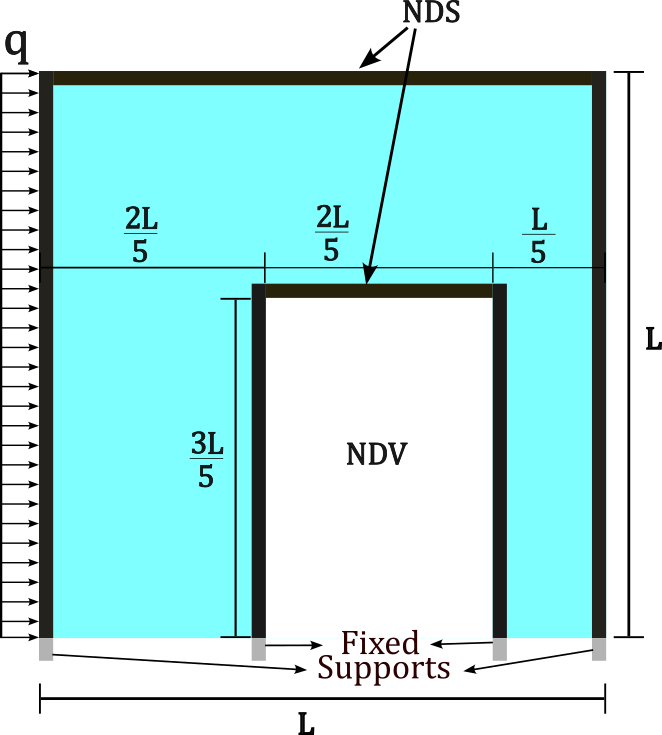}
			\caption{}
			\label{Fig:P2_Schematic}
		\end{subfigure}
		\hfill
		\begin{subfigure}[c]{0.45\textwidth}
			\centering			\includegraphics[height=8cm,width=\textwidth,keepaspectratio]{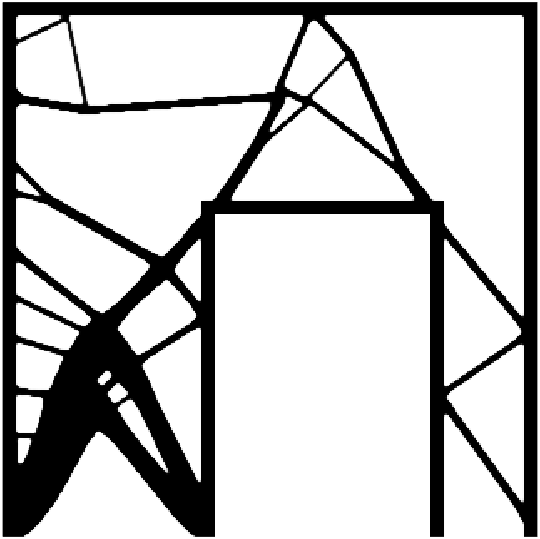}
			\caption{}
			\label{Fig:P2_no_buck_design}
		\end{subfigure}		
		\caption{(a) Design domain of the wall reinforcement problem with force and boundary conditions. (b) Optimized minimum volume design.}
        \label{Fig:P2_Schematic_and_no_buck_design}
	\end{figure}
	\begin{figure}[h!]
		\centering
		\begin{subfigure}[b]{0.45\textwidth}
			\centering
			\includegraphics[width=\textwidth]{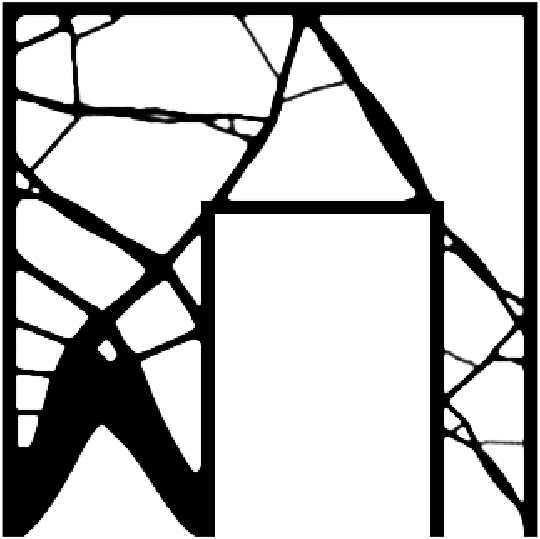} 
			\caption{}
			\label{Fig:P2_conven_method_design}
		\end{subfigure}
		\hfill
		\begin{subfigure}[b]{0.45\textwidth}
			\centering
			\includegraphics[width=\textwidth]{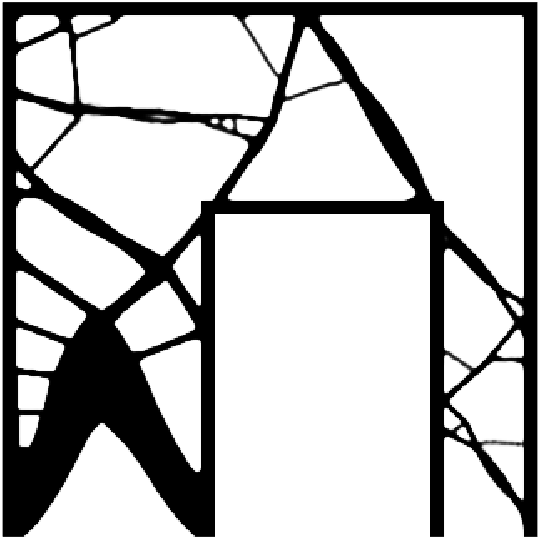} 
			\caption{}
			\label{Fig:P2_ELTD_method_design}
		\end{subfigure}
		\caption{Minimum volume design for wall reinforcement problem obtained using (a) Conventional method and (b) ELTD method}
        \label{Fig:P2_conven_and_ELTD_buck_design}
	\end{figure}

		\begin{figure}
		\begin{subfigure}[t]{0.5\textwidth}
        \centering
			\begin{tikzpicture}
				\begin{axis}[
					xlabel={Iteration},
					ylabel={Objective function $\frac{f(\bm{x})}{f_0}$},
					xmin=0,xmax=600,
					ymin=0.2, ymax=1,
                    width=\linewidth,
					]    
					\addplot[solid,{black}, line width=1pt] table[x index=0, y index=1]
                    {P2_Obj_val_variation_conven_method.txt}; \addlegendentry{Conventional method}
					\addplot[dashed,{red}, line width=1pt, mark = none] table[x index=0, y index=1] {P2_Obj_val_variation_ELTD_method.txt};
					\addlegendentry{ELTD method}
				\end{axis}
			\end{tikzpicture}
			\caption{}
        \label{Fig:P2_Obj_func_variation}
		\end{subfigure}
		\hfill
		\begin{subfigure}[t]{0.5\textwidth}
        \centering
			\begin{tikzpicture}
				\begin{axis}[
					xlabel={Iteration},
					ylabel={Eigenvalues $(\lambda)$},
					xmin=0,xmax=600,
					ymin=0, ymax=1.6,
                    width=\linewidth,
					]    
					\addplot[solid,{red}, line width=1pt] table[x index=0, y index=1] {P2_Eig_val_1_variation_conven_method.txt}; \addlegendentry{$\lambda_1$(Conventional Method)}
                    
					\addplot[solid,{blue}, line width=1pt, mark = none] table[x index=0, y index=1] {P2_Eig_val_2_variation_conven_method.txt};
                    \addlegendentry{$\lambda_2$(Conventional Method)}

                     \addplot[dashed,{black}, line width=1pt, mark = none] table[x index=0, y index=1] {P2_Eig_val_1_variation_ELTD_method.txt};
                     \addlegendentry{$\lambda_1$(ELTD Method)}

                     \addplot[dashed,{green}, line width=1pt, mark = none] table[x index=0, y index=1] {P2_Eig_val_2_variation_ELTD_method.txt};
                     \addlegendentry{$\lambda_2$(ELTD Method)}
                     
				\end{axis}
			\end{tikzpicture}
			\caption{}
        \label{Fig:P2_Eig_val_variation}
		\end{subfigure}
		\caption{Graph showing the evolution of (a) objective function and (b) first two eigenvalues for conventional and ELTD methods for wall Reinforcement problem.} 
	\end{figure}
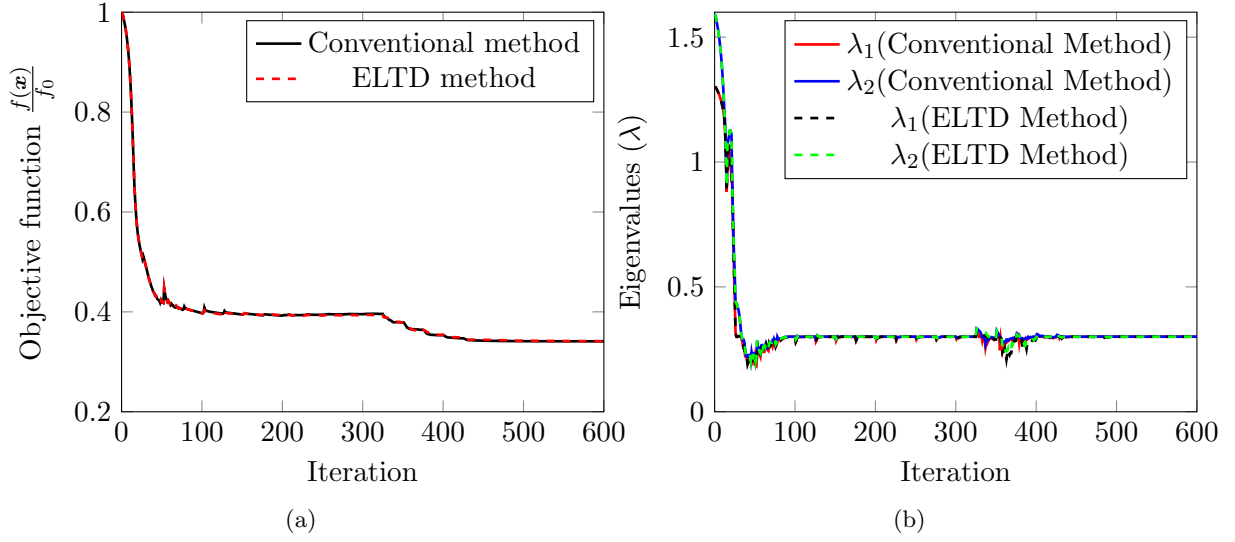

\begin{table}[H]
\centering
\caption{Comparison of the conventional and ELTD approaches for the wall reinforcement problem.}
\label{Tab:P2_final_results}
\renewcommand{\arraystretch}{1.15}
\setlength{\tabcolsep}{5pt}

\resizebox{\textwidth}{!}{%
\begin{tabular}{c| c| c| c| c| c}
\hline\hline
Approach &
Parameter &
Eigenvalues &
$g_C$ &
$f(\bm{x})/f_0$ &
Computational time (s)
\\
\hline

Conventional & $m=12$ &
$\{0.3004,\,0.3007,\,0.3016,$
\newline
$0.3039,\,0.3090\}$ &
$-1.15\times10^{-4}$ &
0.34107 &
6363.88
\\

Conventional & $m=15$ &
$\{0.3004,\,0.3007,\,0.3014,$
\newline
$0.3038,\,0.3082\}$ &
$-8.93\times10^{-5}$ &
0.34151 &
7475.19
\\

ELTD & $\varepsilon=10^{-9}$ &
$\{0.3005,\,0.3009,\,0.3034,$
\newline
$0.3039,\,0.3079\}$ &
$-6.53\times10^{-4}$ &
0.34153 &
6695.79
\\

\hline\hline
\end{tabular}%
}
\end{table}

\subsubsection{Wall reinforcement problem}\label{Sec:Example2ELTD}

The design domain, along with the force and boundary conditions, is depicted in Fig.~\ref{Fig:P2_Schematic}. $L_x=L$ and $L_y= L$, where $L=1$ is set. The thickness $t=1$ is taken. The domain includes non-design solid (NDS) regions with dimensions $ L\times\frac{L}{40}$ along the top and $\frac{L}{40}\times L$ along the left and right edges.  A  non-design void (NDV) region of dimension $\frac{2L}{5} \times \frac{3L}{5}$ is introduced, and is further reinforced by surrounding NDS regions with dimensions $\frac{2L}{5}\times \frac{L}{40}$ along the top and $\frac{L}{40}\times \frac{5L}{8}$ along the left and right edges. The bottom edges of the NDS domains are fixed. A uniformly distributed load $q = \num{1e-2}$ is applied on the left edge in the positive $x-$direction. $320 \times 320$ bilinear FEs are employed to discretize the domain. 

In the first step, optimization problem mentioned in Eq.~\ref{Eq:Axially_Opt_prolem} is solved with $\bar{C} = 2.5C_{\text{loop} = 1}$. The obtained optimized design is depicted in Fig.~\ref{Fig:P2_no_buck_design}, which has a volume fraction of 0.2401 and a critical buckling load of 0.0432. In the next step, the structure's stability is enhanced by imposing a buckling constraint. We solve a minimum volume problem subject to buckling and compliance constraints as:
\begin{equation}
	\begin{aligned}
		&\min_{\bm{x}} \quad \frac{f({\bm{x}})}{{f_0}} \\
		&\text{s.t.} \quad g_\lambda({\bm{x}}) = \overline{\lambda}\times C_\text{KS}[r_i]({\bm{x}}) - 1 \\
        & \quad \quad g_C({\bm{x}}) = C({\bm{x}})/\bar{C} - 1\\
	\end{aligned} 
\end{equation}
	$\overline{\lambda}$ corresponds to the lower bound on the critical buckling load factor, which is taken to be 0.3. 

The results obtained by the conventional method and ELTD method are shown in Table.~\ref{Tab:P2_final_results}. The obtained final values using the ELTD and conventional methods are close to each other. The optimized designs obtained from the conventional and ELTD methods, depicted in Fig.~\ref{Fig:P2_conven_method_design} and Fig.~\ref{Fig:P2_ELTD_method_design}, respectively, also closely resemble each other. The convergence plots for volume fraction and eigenvalues indicated in Fig.~\ref{Fig:P2_Obj_func_variation} and Fig.~\ref{Fig:P2_Eig_val_variation} for both methods follow the same convergence paths. These observations indicate the success of the proposed ELTD method for problems with multiple eigenvalues coalescence. 

The conventional method takes 6363.88 sec, whereas the ELTD approach takes 6695.79 sec (i.e., slightly higher). The required computational time as optimization progresses is mentioned in Fig.~\ref{Fig:P2_Time_variation}. The ELTD method uses MATLAB \texttt{eigs} function as mentioned above (flowchart~\ref{fig:flowchart}) to determine eigenvalues in each iteration.  The proposed ELTD scheme requires nearly the same time as the conventional method, even though the average number of eigenvalues computed per iteration in the former (approximately 6) is less than the number of eigenvalues calculated in the conventional method (12 eigenvalues), as seen in Fig.~\ref{Fig:P2_nEig_agg_variation}. This behavior can be attributed to two main reasons: 
    \begin{itemize}
        \item In a few iterations, changes in the penalization parameters necessitate recalculation of the entire set of eigenvalues.
        \item As eigenvalues become closely spaced, computing even a smaller subset of eigenvalues becomes more computationally demanding. This situation occurs toward the end of the optimization process, where the first 6–8 eigenvalues become clustered.
    \end{itemize}
		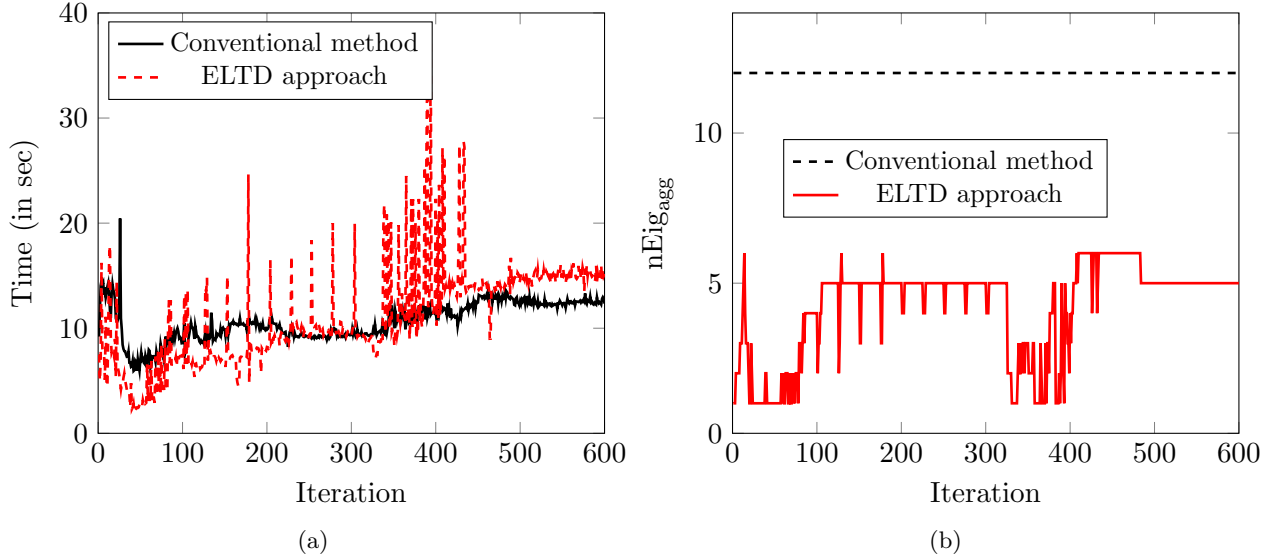
\begin{figure}[htbp]
		\begin{subfigure}[t]{0.52\textwidth}
        \centering
			\begin{tikzpicture}
				\begin{axis}[
					xlabel={Iteration},
					ylabel={Time (in sec)},
					xmin=0,xmax=600,
					ymin=0, ymax=40,
                    width=\linewidth,
                    legend style={ at={(0.02,0.98)},
                    anchor = north west,
                    font=\small}
					]    
					\addplot[solid,{black}, line width=1 pt] table[x index=0, y index=1] {P2_Time_variation_conven_method.txt}; \addlegendentry{Conventional method}

                    \addplot[dashed,{red}, line width=1 pt, mark = none] table[x index=0, y index=1] {P2_Time_variation_ELTD_method.txt};
                    \addlegendentry{ELTD approach}   
				\end{axis}
			\end{tikzpicture}
			\caption{}
        \label{Fig:P2_Time_variation}
		\end{subfigure}
        \hfill
        \begin{subfigure}[t]{0.52\textwidth}
        \centering
			\begin{tikzpicture}
				\begin{axis}[
					xlabel={Iteration},
					ylabel={$\text{nEig}_\text{agg}$},
					xmin=0,xmax=600,
					ymin=0, ymax=14,
                    width=\linewidth,
                    legend style={ at={(0.1,0.7)},
                    anchor=north west,
                    font=\small}
					]    
					\addplot[dashed,{black}, line width=1 pt] table[x index=0, y index=1] {P2_nEig_agg_variation_conven_method.txt}; \addlegendentry{Conventional method}
					\addplot[solid,{red}, line width= 1 pt, mark = none] table[x index=0, y index=1] {P2_nEig_agg_variation_ELTD_method.txt};
					\addlegendentry{ELTD approach}
				\end{axis}
			\end{tikzpicture}
			\caption{}
        \label{Fig:P2_nEig_agg_variation}
		\end{subfigure}
		\caption{Graph showing (a) Time taken to calculate eigenvalues and eigenvectors in a given iteration and (b) Number of eigenvalues aggregated ($\text{nEig}_\text{agg}$)in each iteration in Conventional and ELTD methods for Wall Reinforcement problem. The large spike in the Time (vs) Iteration graph corresponds to a case in which the entire set of eigenvalues is recalculated.} 
        \label{Fig:P2_Time_and_nEig_variation}
	\end{figure}
	
	As illustrated in Table~\ref{Tab:P2_final_results}, the proposed ELTD scheme takes slightly longer to solve this problem when multiple eigenvalues coalesce. Ref.~\cite{ferrari2021topology} takes 12 eigenvalues to begin with. However, one typically uses a larger number of eigenvalues ($m\in(30,\,50)$) for a given problem~\cite{torii2017structural,ferrari2019revisiting,ferrari2021topology,chin2016large}. When we aggregate 15 eigenvalues instead of 12 across all iterations for the conventional method, the required computational time is 7475.19 sec, which is more than that required by the ELTD approach (Table~\ref{Tab:P2_final_results}). Therefore, even with only 3 additional eigenvalues, the computational time exceeds that of the ELTD scheme. 

The computational cost of evaluating the prescribed number of eigenvalues with the \texttt{eigs} function depends on several factors, including the choice of initial guess, memory effects, and multi-threading. In the present study, a uniform initial guess vector with all entries set to 1 is used at each iteration to ensure consistency. Moreover, all computations are performed in a single-threaded environment to maintain a fair comparison of computational performance. Each problem is solved three times, and the reported computational time is the average, with a standard deviation of approximately 30 seconds. To further enhance the reliability of the computed spectrum, four additional eigenvalues beyond the required number are evaluated, ensuring the accuracy of the highest computed modes, as recommended in~\cite{ferrari2021topology}.


\subsubsection{Shear loaded hinged plate}\label{Sec:Example3ELTD}
\begin{figure}[h!]
		\centering
		\begin{subfigure}[c]{0.49\textwidth}
			\centering	\includegraphics[height=8cm,width=\textwidth,keepaspectratio]{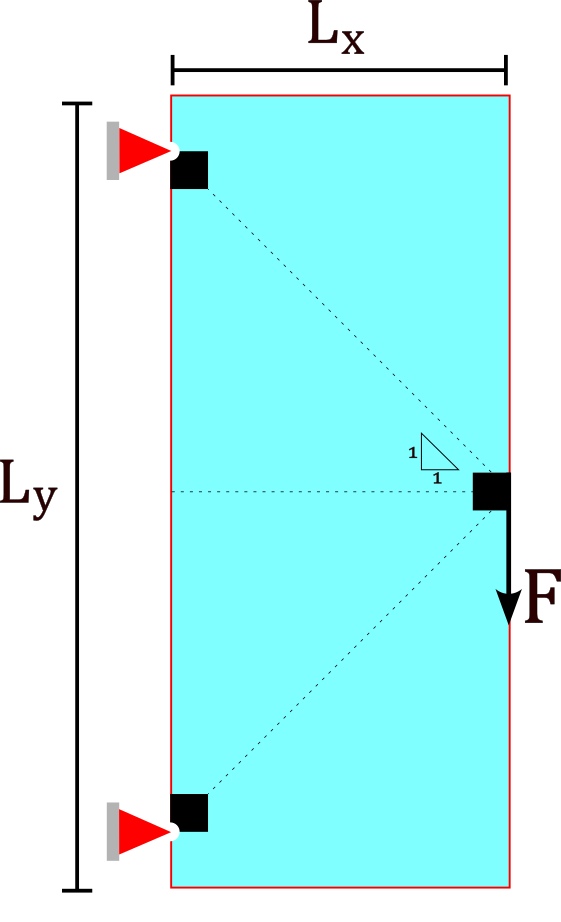}
			\caption{}
			\label{Fig:P3_Schematic}
		\end{subfigure}
		\hfill
		\begin{subfigure}[c]{0.49\textwidth}
			\centering			\includegraphics[height=8cm,width=\textwidth,keepaspectratio]{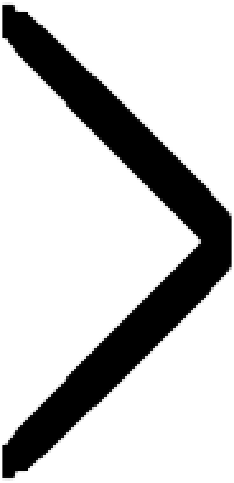}
			\caption{}
			\label{Fig:P3_no_buck_design}
		\end{subfigure}		
		\caption{(a) Design domain of the shear-loaded hinged plate problem with force and boundary conditions. (b) Optimized minimum compliance design.}
        \label{Fig:P3_Schematic_and_no_buck_design}
	\end{figure}

A 2D plate hinged at two points with $L_{x} = 1$ and $L_{y} = 7/3$ is depicted in Fig.~\ref{Fig:P3_Schematic}, which is solved here. The domain is discretized by $90 \times 210$ bilinear elements. The domain includes NDS regions spanning over $\frac{L_x}{90}\times\frac{L_y}{21}$ FEs near the right edge and near both hinges. A shear force of magnitude $F = \num{2e-3}$ is distributed over the right edge of the NDS region. The optimization problem is framed as:
 \begin{equation}\label{Eq:Shear_load_prolem}
        \begin{aligned}
		&\min_{\bm{x}} \quad C({\bm{x}})/{C_0} \\
		&\text{s.t.} \quad g_\text{V}({\bm{x}}) = f({\bm{x}})/\bar{f} - 1 \leq 0
	\end{aligned}
    \end{equation}
Here $f({\bm{x}})$, $c_0$ and  $\bar{f}$ represent the volume of the structure, the compliance of the initial guess and the allowed volume, which is equal to 0.2 $\times$ the volume of the total structure, respectively. 

\begin{table}[H]
\centering
\caption{Comparison of the conventional and ELTD approaches for the shear-loaded hinged plate problem.}
\label{Tab:P3_final_results}
\renewcommand{\arraystretch}{1.15}
\setlength{\tabcolsep}{5pt}

\resizebox{\textwidth}{!}{%
\begin{tabular}{c| c| c| c| c| c}
\hline\hline
Approach &
Parameter &
Eigenvalues &
$g_V$ &
$C(\bm{x})/c_0$ &
Computational time (s)
\\
\hline

Conventional & $m=24$ &
$\{0.7504,\,0.7655,\,0.7817,$
\newline
$0.7920,\,0.8221,\,0.8449,$
\newline
$0.8463\}$ &
$-1.05\times10^{-7}$ &
1.6163 &
4341.62
\\

ELTD & $\varepsilon=10^{-9}$ &
$\{0.7503,\,0.7676,\,0.7899,$
\newline
$0.7914,\,0.8042,\,0.8189,$
\newline
$0.8465\}$ &
$-1.35\times10^{-6}$ &
1.5858 &
2371.48
\\

\hline\hline
\end{tabular}%
}
\end{table}

	
	\begin{figure}[h!]
		\centering
		\begin{subfigure}[b]{0.45\textwidth}
			\centering			\includegraphics[width=0.45\textwidth,height=8cm]{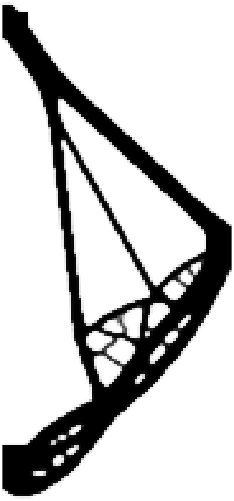} 
			\caption{}
			\label{Fig:P3_conven_method_design}
		\end{subfigure}
		\hfill
		\begin{subfigure}[b]{0.45\textwidth}
			\centering			\includegraphics[width=0.45\textwidth,height=8cm]{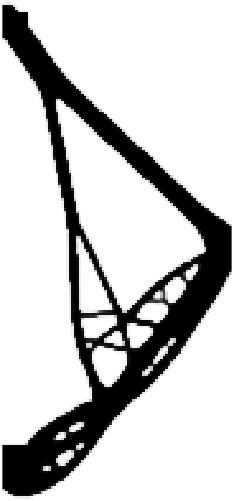} 
			\caption{}
			\label{Fig:P3_ELTD_method_design}
		\end{subfigure}
		\caption{Minimum compliance design of shear-loaded hinged plate obtained using (a) Conventional method and (b) ELTD method}
        \label{Fig:P3_conven_and_ELTD_buck_design}
	\end{figure}

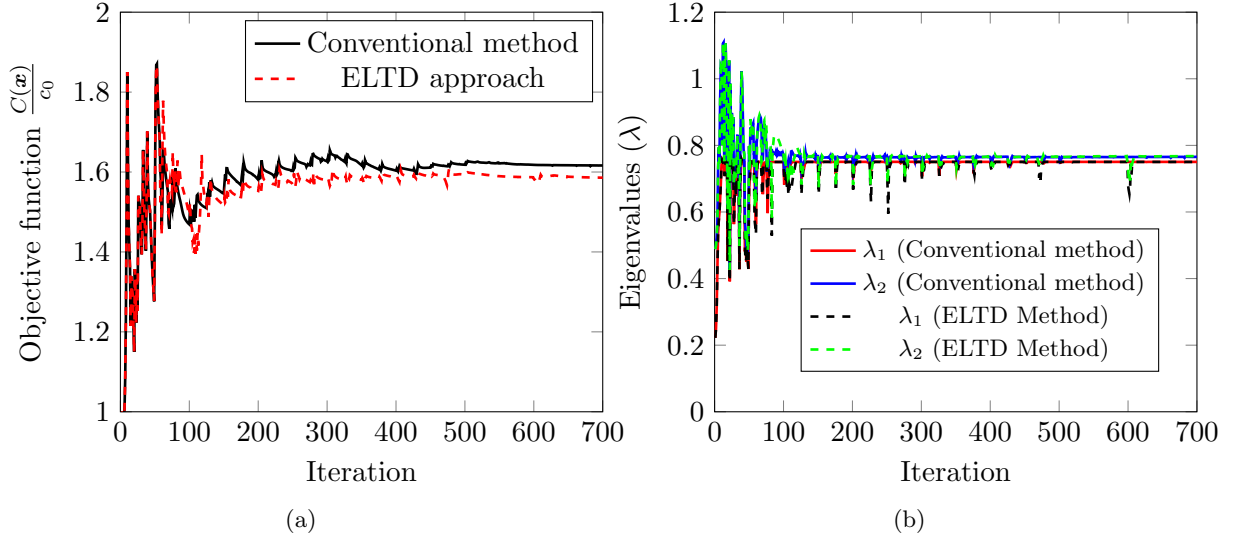
\begin{figure}[H]
		\begin{subfigure}[t]{0.5\textwidth}
        \centering
			\begin{tikzpicture}
				\begin{axis}[
					xlabel={Iteration},
					ylabel={Objective function $\frac{C(\bm{x})}{c_0}$},
					xmin=0,xmax=700,
					ymin=1, ymax=2,
                    width=\linewidth,
					]    
					\addplot[solid,{black}, line width=1pt] table[x index=0, y index=1] {P3_Obj_val_variation_conven_method.txt}; \addlegendentry{Conventional method}
					\addplot[dashed,{red}, line width=1pt, mark = none] table[x index=0, y index=1] {P3_Obj_val_variation_ELTD_method.txt};
					\addlegendentry{ELTD approach}
				\end{axis}
			\end{tikzpicture}
			\caption{}
        \label{Fig:P3_Obj_val_variation}
		\end{subfigure}
		\hfill
		\begin{subfigure}[t]{0.5\textwidth}
        \centering
			\begin{tikzpicture}
				\begin{axis}[
					xlabel={Iteration},
					ylabel={Eigenvalues $(\lambda)$},
					xmin=0,xmax=700,
					ymin=0, ymax=1.2,
                    width=\linewidth,
                    legend style={
                    at={(0.55,0.1)},
                    anchor=south,
                    font=\footnotesize},
					]    
					\addplot[solid,{red}, line width=1pt] table[x index=0, y index=1] {P3_Eig_val_1_variation_conven_method.txt}; \addlegendentry{$\lambda_1$ (Conventional method)}
                    
					\addplot[solid,{blue}, line width=1pt, mark = none] table[x index=0, y index=1] {P3_Eig_val_2_variation_conven_method.txt};
                    \addlegendentry{$\lambda_2$ (Conventional method)}

                     \addplot[dashed,{black}, line width=1pt, mark = none] table[x index=0, y index=1] {P3_Eig_val_1_variation_ELTD_method.txt};
                     \addlegendentry{$\lambda_1$ (ELTD Method)}

                     \addplot[dashed,{green}, line width=1pt, mark = none] table[x index=0, y index=1] {P3_Eig_val_2_variation_ELTD_method.txt};
                     \addlegendentry{$\lambda_2$ (ELTD Method)}

				\end{axis}
			\end{tikzpicture}
			\caption{}
        \label{Fig:P3_Eig_val_variation}
		\end{subfigure}
		\caption{The graph showing the evolution of (a) the objective function and (b) the first two eigenvalues for both methods in the shear-loaded hinged plate problem.}
        \label{Fig:P3_Obj_and_Eig_val_variation}
	\end{figure}

     The optimized design obtained is depicted in Fig.~\ref{Fig:P3_no_buck_design}, with a normalized compliance of 0.0774 and a critical buckling load of 0.1974. In the next step, the structure's stability is enhanced by imposing a buckling constraint. We solve a minimum compliance problem subject to buckling and volume constraints as
\begin{equation}
	\begin{aligned}
		&\min_{\bm{x}} \quad C({\bm{x}})/{C_0} \\
		&\text{s.t.} \quad g_\lambda({\bm{x}}) = \overline{\lambda}\times C_\text{KS}[r_i]({\bm{x}}) - 1 \\
        & \quad \quad g_V({\bm{x}}) = f({\bm{x}})/\bar{f} - 1
	\end{aligned} 
    \label{Eq:opt_formulation_prob_3}
\end{equation}
	 The value of $\overline{\lambda}$ is taken to be 0.75. Since the lower bound in the buckling constraint is high, there is a possibility that a large number of eigenvalues are coming closer. Thus, 24 eigenvalues are aggregated as per \cite{ferrari2019revisiting}. The final results of the conventional and ELTD approaches are summarized in Table.~\ref{Tab:P3_final_results}, and the optimized designs obtained from the conventional and ELTD methods are depicted in Fig.~\ref{Fig:P3_conven_method_design} and Fig.~\ref{Fig:P3_ELTD_method_design}, respectively. Both designs look broadly similar, but there are certain differences in how the thin structures are aligned. As mentioned above, the conventional method aggregates 24 eigenvalues in each iteration. However, ideally, only very few eigenvalues are relevant in many iterations. This will make the aggregation function overly conservative. These small differences, compounded over many iterations, can lead to different final designs between conventional and ELTD approaches. We can also observe this from the convergence plots of compliance and eigenvalues as shown in Fig.~\ref{Fig:P3_Obj_val_variation} and Fig.~\ref{Fig:P3_Eig_val_variation} for both methods. The objective function predicted by the ELTD method is consistently smaller than that of the conventional method from the $100$-th iteration onwards.


		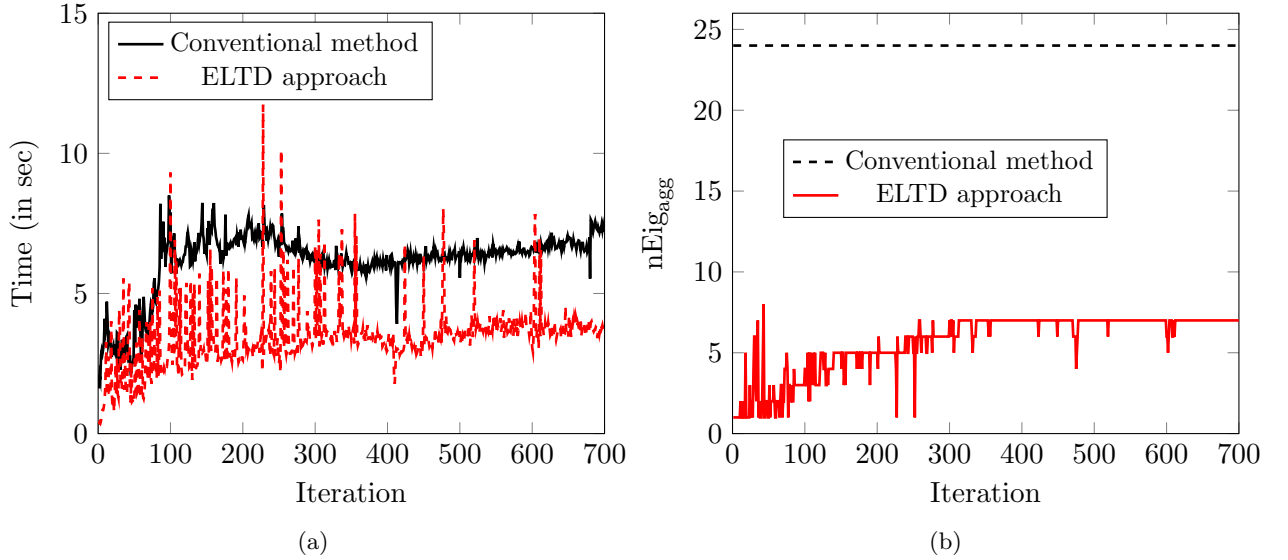
\begin{figure}[htbp]
		\begin{subfigure}[t]{0.52\textwidth}
        \centering
			\begin{tikzpicture}
				\begin{axis}[
					xlabel={Iteration},
					ylabel={Time (in sec)},
					xmin=0,xmax=700,
					ymin=0, ymax=15,
                    width=\linewidth,
                    legend style={ at={(0.02,0.98)},
                    anchor = north west,
                    font=\small}
					]    
					\addplot[solid,{black}, line width=1 pt] table[x index=0, y index=1] {P3_Time_variation_conven_method.txt}; \addlegendentry{Conventional method}

                    \addplot[dashed,{red}, line width=1 pt, mark = none] table[x index=0, y index=1] {P3_Time_variation_ELTD_method.txt};
                    \addlegendentry{ELTD approach}   
				\end{axis}
			\end{tikzpicture}
			\caption{}
        \label{Fig:P3_Time_variation}
		\end{subfigure}
        \hfill
        \begin{subfigure}[t]{0.52\textwidth}
        \centering
			\begin{tikzpicture}
				\begin{axis}[
					xlabel={Iteration},
					ylabel={$\text{nEig}_\text{agg}$},
					xmin=0,xmax=700,
					ymin=0, ymax=26,
                    width=\linewidth,
                    legend style={ at={(0.1,0.7)},
                    anchor=north west,
                    font=\small}
					]    
					\addplot[dashed,{black}, line width=1 pt] table[x index=0, y index=1] {P3_nEig_agg_variation_conven_method.txt}; \addlegendentry{Conventional method}
					\addplot[solid,{red}, line width= 1 pt, mark = none] table[x index=0, y index=1] {P3_nEig_agg_variation_ELTD_method.txt};
					\addlegendentry{ELTD approach}
				\end{axis}
			\end{tikzpicture}
			\caption{}
        \label{Fig:P3_nEig_agg_variation}
		\end{subfigure}
		\caption{Graph showing (a) Time taken to calculate eigenvalues and eigenvectors in a given iteration and (b) Number of eigenvalues aggregated ($\text{nEig}_\text{agg}$)in each iteration in Conventional and ELTD methods for shear-loaded hinged plate problem. The large spike in the Time (vs) Iteration graph corresponds to a case in which the entire set of eigenvalues is recalculated.} 
        \label{Fig:P3_Time_and_nEig_variation}
	\end{figure}
The conventional method takes 4341.62 sec, whereas the ELTD approach takes 2371.48 sec (Table.~\ref{Tab:P3_final_results}). There is a 45.37\% reduction in computational time for this problem. This is because only a small number of eigenvalues are relevant towards the end of the optimization process, which is elucidated in Fig.~\ref{Fig:P3_nEig_agg_variation}.
  
 We test the proposed ELTD approach on three problems: a compressed column, wall reinforcement, and a shear-loaded hinged plate. In these problems, buckling load, compliance, and structural volume serve as the objectives and constraints. In all three cases, the optimized geometry is nearly identical to that obtained by the conventional method. However, the number of relevant eigenvalues needed during optimization is much smaller than the number typically included. Including only the relevant eigenvalues makes the buckling constraint less overly conservative, allowing the algorithm to get much closer to the true multimodal optimum. In fact, ELTD achieved better (lower) objective values than the traditional approach across all test cases. Next, we address the third issue: determining the optimal solution's modality for buckling problems.


\section{Modality at the optimum solution}\label{Sec4:ModatOpt}

As noted earlier (Fig.~\ref{fig:multimodal limits}), the aggregation functions do not provide an exact optimal point for finite values of $\rho$ (Eq.~\ref{Eq:KS_function}) or $p$ (Eq.~\ref{Eq:P_function}). For the $m$-modal eigenvalue $\left( r_i = r, \forall \hspace{0.5em} i = 1,2,3,\,..,m \right)$, Eq.~\ref{Eq:KS_der} and Eq.~\ref{Eq:pnorm_der} at optimum point yield to

	\begin{equation}
		\nabla C_{\mathrm{KS}}(r_i, \rho) = \frac{-\sum_{i=1}^{m} \bm{\phi}_i^{\top} \left( \nabla \mathbf{G} + r \, \nabla \mathbf{K} \right) \bm{\phi}_i}{m} = 0
	\end{equation}
	\begin{equation}
		\nabla c_p\left(r_i, p\right)=\frac{-\sum_{i=1}^m \bm{\phi_i}^{\top}(\nabla \mathbf{G}+r \nabla \bm{K}) \bm{\phi_i}}{m^{1 - \frac{1}{p}}}=0  
	\end{equation}
    Optimality conditions indicated by the above equations are not the same as those provided in Seyranian et al.\cite{seyranian1994multiple}. Additionally, since the considered problem is $ m$-modal, any orthonormal basis of the associated eigenspace can be used in the formulation. Thus, aggregation functions yield solutions close to the point of eigenvalue coalescence rather than the true optimum, indicating that the problem is unimodal. However, for very high parameter ($\rho \hspace{0.2em}  \text{or}\hspace{0.2em}  p$), the optimum solution shows the sign of multimodality. If the selected $\rho$ is below the required threshold, the aggregation may fail to reveal multimodality, leading to the incorrect assumption that the problem is unimodal. In such cases, the evaluated eigenvalues are not sufficiently close to indicate potential coalescence at the optimum. Consequently, the associated eigenspace cannot be characterized, as the number of eigenvectors required to construct its linear span remains unknown. This limitation becomes critical when interpreting physical responses, such as the displacement field of a buckled configuration, which requires accurate information about the underlying eigenvectors.
    
   To the best of authors' knowledge, at present, no reliable and robust method exists to predict the modality of eigenvalues at a given point correctly. Typically, the relative difference between consecutive eigenvalues is used, but this approach is unreliable because it is problem-dependent~\cite{seyranian1994multiple}. In practice, using a tolerance (such as $\num{1e-4}$) to check relative differences often requires a relaxation to reduce errors arising from finite aggregation parameters.  Additionally, aggregating more eigenvalues than necessary can degrade the approximation of $\max(r_i)$, potentially leading to incorrect modality. The optimization algorithm may also oscillate near the true optimum, preventing proper convergence. As a result, the choice of convergence criteria directly affects the perceived modality. The classic example below illustrates these issues.
   \begin{figure}[htbp]
    \centering

    \begin{subfigure}[c]{0.45\textwidth}
        \centering
        \includegraphics[width=\linewidth]{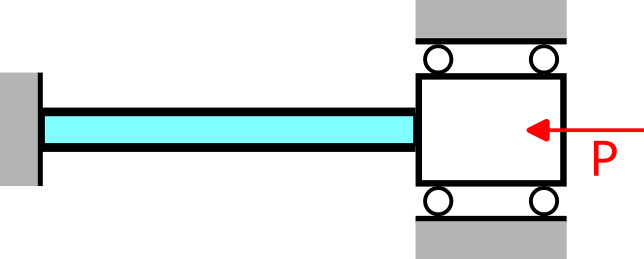}
        \caption[b]{}
        \label{Fig:P4_Schematic}
    \end{subfigure}
    \begin{subfigure}[c]{0.48\textwidth}
        \centering
        \begin{tikzpicture}
            \begin{axis}[
                width=\linewidth,
                xlabel={$x$},
                ylabel={$H(x)$},
                xmin=0,
                xmax=1,
                ymin=-0.6,
                ymax=0.6
            ]

                \addplot[blue,solid,line width=1pt]
                table[x index=0,y index=1]{P4_LBFGSB_area_variation_bottom_half.txt};

                \addplot[blue,solid,line width=1pt]
                table[x index=0,y index=1]{P4_LBFGSB_area_variation_top_half.txt};

                \addplot[black,dashed,line width=1pt]
                table[x index=0,y index=1]{P4_LBFGSB_area_variation_reference_line.txt};

            \end{axis}
        \end{tikzpicture}
        \caption{}
        \label{Fig:P4_LBFGSB_area_variation}
    \end{subfigure}

    \caption{(a) Schematic of a fixed-fixed 1D column  subjected to a compressive load. (b) Optimum column shape $H(x)=\frac{1}{2}\sqrt{A(x)}$ that maximizes the critical buckling load according to Eq.~22.}
    \label{Fig:P4_Schematic_and_LBFGSB_buck_design}
\end{figure}
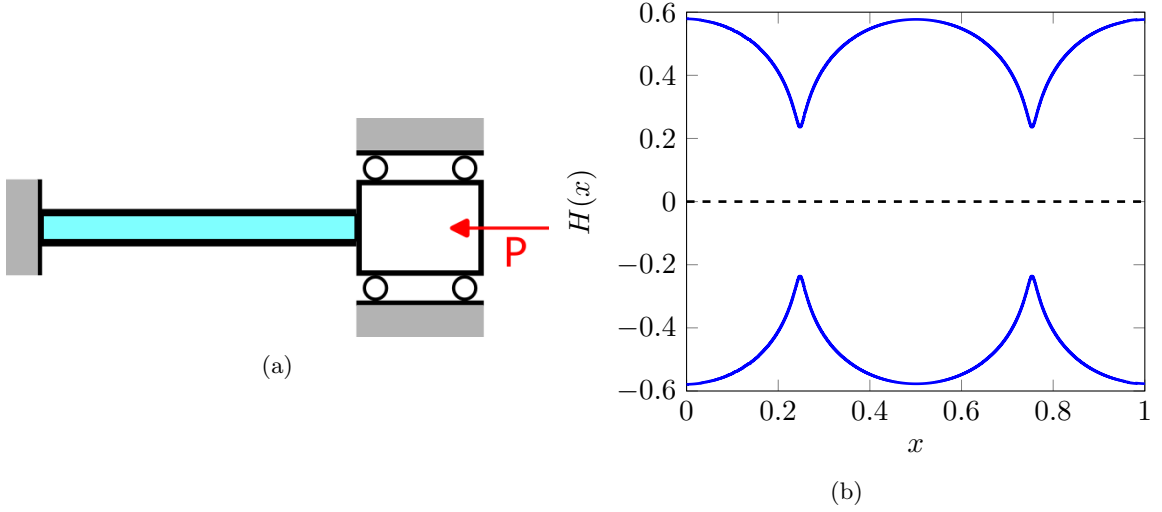

\begin{table}[H]
\centering
\caption{Mesh convergence study based on the first two eigenvalues and the volume constraint. Convergence is achieved when the $L_{\infty}$ norm of the change in design variables is less than $10^{-4}$.}
\label{Tab:P4_parametric_study_num_elements}
\renewcommand{\arraystretch}{1.15}
\setlength{\tabcolsep}{8pt}

\begin{tabular}{c| c| c| c| c| c}
\hline\hline
No. of elements &
$\lambda_1$ &
$\lambda_2$ &
$\lambda_2-\lambda_1$ &
Relative difference &
$g_V$
\\
\hline

100   & 4.3516 & 4.4650 & 0.1134 & 0.0261 & $-1.22\times10^{-7}$ \\
500   & 4.3573 & 4.4673 & 0.1100 & 0.0252 & $-1.36\times10^{-6}$ \\
1000  & 4.3575 & 4.4677 & 0.1102 & 0.0253 & $-9.89\times10^{-8}$ \\
2500  & 4.3575 & 4.4673 & 0.1098 & 0.0252 & $-7.10\times10^{-8}$ \\
5000  & 4.3575 & 4.4672 & 0.1097 & 0.0252 & $-7.08\times10^{-8}$ \\
7500  & 4.3580 & 4.4681 & 0.1101 & 0.0253 & $-7.84\times10^{-8}$ \\
10000 & 4.3583 & 4.4666 & 0.1083 & 0.0248 & $-7.01\times10^{-7}$ \\

\hline\hline
\end{tabular}
\end{table}


\subsection{Fixed-fixed 1D column problem and challenges with aggregation method}\label{Sec:fixfix1DColume}
Consider a 1D column with a square cross-section and fixed at both ends as shown in Fig.~\ref{Fig:P4_Schematic_and_LBFGSB_buck_design}a. The length of the column ($l$) is 1 meter, and its Young's modulus $(E)$ is 1 Pa. The area moment of inertia is $\alpha A^2$ where $\alpha=\frac{1}{12}$. We aim to find the optimal cross-sectional area as a function of length, $A(x)$, to maximize the critical buckling load, subject to a maximum volume limit $V_0 = 1\text{ m}^3$. In the parameterized setting, we use 1000 1D bar FEs to discretize the column. The optimization problem is written for the column problem as:
	\begin{equation}
	\begin{aligned}
		&\min \quad C_\text{KS}[r_i]({\mathbf{A}}) \\
		&\text{s.t.} \quad g_V(\mathbf{A}) = \frac{\sum_{i=1}^{1000} A_il_i}{V_0} - 1 \leq 0
	\end{aligned}
	\end{equation}
	where ${\mathbf{A}} = [A_1,\, A_2,\cdots,\,A_{1000}]^\top$. $A_i$ and $l_i$ represent the area and length of $i$-th element, respectively. 

The analytical solution of the problem is well-documented in Refs.~\cite{olhoff1977single,masur1984optimal,bratus1983bimodal}. The critical buckling load of the optimized column is $P = \frac{52.3563E\alpha V_{0}^2}{L^4} = 4.363025$ with bimodality and the optimal column shape is illustrated in Fig.~\ref{Fig:P4_LBFGSB_area_variation}. When the KS-aggregation function with an aggregation parameter of 500 and $ m = 2$ is employed to solve the problem, we get  $\lambda_1 = 4.3575$ and $\lambda_2 = 4.4677$, i.e., $(\lambda_2 -\lambda_1) = 0.1102$ or relative difference  $(\lambda_2 -\lambda_1)/(\lambda_1) = 0.0253$. We increase the number of FEs to evaluate its impact, but the solution remains the same beyond 1,000 elements, as shown in Table~\ref{Tab:P4_parametric_study_num_elements}. According to the analytical solution~\cite{olhoff1977single}, the column problem (Fig.~\ref{Fig:P4_Schematic_and_LBFGSB_buck_design}) is bimodal. However, the numerical results using the aggregation method in Table~\ref{Tab:P4_parametric_study_num_elements} do not clearly reflect this. $\lambda_1$ and $\lambda_2$ come close to each other; however, it does not mean the eigenvalues coalesce. Therefore, we need a robust approach to determine how many eigenvalues are actually coalescing at the optimum point. 

\begin{figure}
\centering
\begin{tikzpicture}
\begin{axis}[
    name=main,
    width=10cm,
    height=7cm,
    xlabel={$x$},
    ylabel={$\lambda$},
    domain=-4:4,
    samples=200,
    grid=both,
    legend pos=north west
]
\addplot[thick, solid] {-x^2 + 4};
\addlegendentry{$\lambda_1 = -x^2 + 4$}

\addplot[thick, dashed] {x^2 + 4*(1 + 1e-4)};
\addlegendentry{$\lambda_2 = x^2 + 4(1 + 10^{-4})$}

\draw[red, thick] (axis cs:-0.1,3.9997) rectangle (axis cs:0.1,4.0005);

\end{axis}

\begin{axis}[
    at={(main.north east)},
    anchor=north west,
    xshift=1cm,
    width=6cm,
    height=4.5cm,
    domain=-0.1:0.1,
    samples=200,
    ymin=3.995,
    ymax=4.01,
    grid=both,
    title={Zoomed view},
    xlabel={$x$},
    ylabel={$\lambda$}
]
\addplot[thick, solid] {-x^2 + 4};
\addplot[thick, dashed] {x^2 + 4*(1 + 1e-4)};
\end{axis}
\end{tikzpicture}
\caption{A plot of $\lambda_1(x) \text{ and } \lambda_2(x)$ showing fake coalescence}\label{Fig:Fake coalescence}
\end{figure}
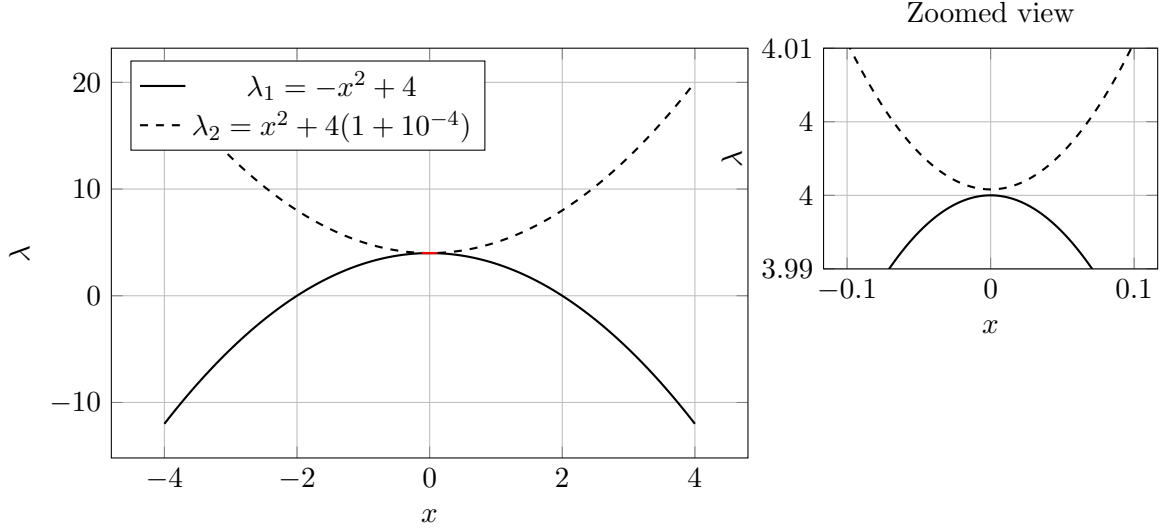
\subsection{A typical example for fake coalescence} \label{subsub:Fakecol}
We illustrate a false coalescence state that arises from relying on relative differences using the following example. Consider $\lambda_1 = -x^2 + 4$ and $\lambda_2 = x^2 + 4(1 + 10^{-4})$. At $x=0$ (Fig.~\ref{Fig:Fake coalescence}), the relative difference between $\lambda_1$ and $\lambda_2$ is $10^{-4}$, yet the eigenvalues do not coalesce. In fact, $x=0$ remains the optimum point even without including $\lambda_2$ in the aggregation function. This shows that checking relative differences does not always identify the true modality.

Another way to locate the coalescence point is to evaluate the directional derivatives of the repeated eigenvalues at the multimodal optimum, where these derivatives should not all have the same sign. In the presented case above (Fig.~\ref{Fig:Fake coalescence}), close to $x=0$, indeed the directional derivatives of $\lambda_1(x)$ and $\lambda_2(x)$ have opposite signs. However, $x=0$ is not the multimodal point. The directional derivative check for the multimodality should only be used when the eigenvalues are exactly coalescing, which is not the case here. This indicates that, in such cases, the directional derivative check can also be misleading when predicting the modality. Therefore, directional derivative tests may also fail and may not provide exactly how many eigenvalues are coalescing for the optimum point obtained numerically.
    
In view of the above discussions and the example presented (Fig.~\ref{Fig:Fake coalescence}), the only method to predict the modality is to approach the actual optimum point by increasing the aggregation parameter using a continuation scheme. For example, for the 1D compressed column problem, to obtain a relative difference of $\num{1e-4}$ between the first two eigenvalues, we need to gradually increase the aggregation parameter to at least
$\num{1e5}$. This makes the analysis computationally very expensive, and we still can't theoretically guarantee the coalescence. To address this gap, we present a new strategy to address coalescence, yielding a very small relative difference on the order of $\num{1e-10}$ with less computational effort. 

\subsection{A new strategy for predicting modality at the optimum point}\label{SubSec4}
In the section above, we note that using a finite aggregation parameter can yield eigenvalues with large relative differences, making it difficult to draw conclusions about the modal behavior. We also demonstrate that small relative differences may sometimes give a misleading indication of eigenvalue coalescence. To address these issues, we propose a new strategy in which a modified (unconstrained) optimization problem is formulated to directly minimize the difference between eigenvalues with additional constraint terms. We first apply this framework to the 1-D column problem to illustrate the key steps and the success of the proposed approach. The primary purpose of determining modality is to visualize and analyze all possible buckling mode shapes.

\subsubsection{Modality of fixed-fixed 1D compressed column}\label{sec:mod1Dcompressed}

The proposed optimization problem to determine eigenvalues coalescence for the 1D column problem presented above in Sec.~\ref{Sec:fixfix1DColume} is written as:
    
\begin{equation}
    \begin{split}
	    \min_{\mathbf{A}} \quad & \left( \sum_{i=1}^{n_e - 1}\sum_{j= i+1}^{n_e} \left(\frac{r_i - r_j}{r_0}\right)^2 \right)+ \left(g_V(\mathbf{A})\right)^2 = \left((n_e - 1)\left(\frac{S_1}{r_0}\right)^2 - 2n_e\frac{S_2}{r_0^2}\right) + \left(g_V(\mathbf{A})\right)^2 \\
        & \text{where} \quad S_1 =  \sum_{i=1}^{n_e} r_i \quad \text{and} \quad S_2 = \sum_{i=1}^{n_e - 1}\sum_{j= i+1}^{n_e} r_ir_j
	\end{split}
    \label{Eq:1Dcolumcolopt}
\end{equation}
We solve the unconstrained optimization problem (Eq.~\ref{Eq:1Dcolumcolopt}) metioned above using the L-BFGS-B algorithm \cite{byrd1995limited,zhu1997algorithm,morales2011remark}. The algorithm's code is taken from the L-BFGS-B-C master repository on GitHub, a popular, modern implementation by Stephen Becker~\cite{becker2023lbfgsbc}. To obtain very small relative differences, we set the tolerances in the termination test to zero. Also, the number of memory vectors used to approximate the inverse-Hessian vectors is set to 10~\cite{byrd1995limited,zhu1997algorithm,morales2011remark}. The maximum number of iterations is set to be 5000. The initial design obtained from the ELTD approach is chosen as the initial guess. The objective function is appropriately normalized so that its value equals 100 in the first iteration.  $r_0$ is set to $r_{max} = r_1$ of the first iteration for normalization purposes. At the initial stage, the number of eigenvalues considered, $n_e$, is set to the same value used in the final iteration of the ELTD method. The value of $n_e$ is then gradually reduced until a sufficiently small relative difference between the eigenvalues is obtained. It is worth noting that, although the individual eigenvalues are not Fr\'echet differentiable, the quantity $\left( \sum_{i=1}^{n_e}\sum_{j=i+1}^{n_e} \left(r_i-r_j\right)^2 \right)$ is Fr\'echet differentiable, as it can be expressed in terms of suitable combinations of symmetric polynomials, as shown in Eq.~\ref{Eq:1Dcolumcolopt}.

In view of the above discussion, we set $n_e=2$. By solving the optimization problem given in Eq.~\ref{Eq:1Dcolumcolopt} using the proposed method, we obtain $\lambda_1=\lambda_2=4.362879$, with a relative difference of only $4.40\times10^{-8}$ after just 17 iterations. This result confirms the bimodal nature of the optimal solution, consistent with the findings reported in~\cite{olhoff1977single}. Note that such a small relative difference is difficult to achieve using the aggregation function approach unless the aggregation parameter is chosen to be very large, typically of the order of $10^{9}$. However, using such large values of $\rho$ generally requires a continuation scheme in which $\rho$ is increased gradually, leading to a significantly larger number of iterations; thus, computational time.
		\begin{figure}[htbp]
		\begin{subfigure}[t]{0.52\textwidth}
        \centering
			\begin{tikzpicture}
				\begin{axis}[
					xlabel={$x$},
					ylabel={$\phi(x)$},
					xmin=0,xmax=1,
					ymin=-0.2, ymax=1,
                    width=\linewidth,
                    legend style={at = {(0.32,0.1)},
                    anchor = south west,
                    font=\small}
					]    
					\addplot[dashed,{red}, line width=1 pt] table[x index=0, y index=1] {P4_buck_mode_minus_phi_1_conven_method.txt}; \addlegendentry{$\bm{-\phi_1(x)}$}
					\addplot[solid,{black}, line width= 1 pt, mark = none] table[x index=0, y index=1] {P4_buck_mode_phi_2_minus_phi_1_conven_method.txt};
					\addlegendentry{$\bm{\phi_2(x) - \phi_1(x)}$}
				\end{axis}
			\end{tikzpicture}
			\caption{}
       \label{Fig:P4_buck_mode_conven_method}
		\end{subfigure}
		\begin{subfigure}[t]{0.52\textwidth}
        \centering
			\begin{tikzpicture}
				\begin{axis}[
					xlabel={$x$},
					ylabel={$\phi(x)$},
					xmin=0,xmax=1,
					ymin=0, ymax=1,
                    width=\linewidth,
                    legend style={ at={(0.26,0.1)},
                    anchor = south west,
                    font=\small}
					]    
					\addplot[dashed,{red}, line width=1 pt] table[x index=0, y index=1] {P4_buck_mode_minus_phi_1_LBFGSB_method.txt}; \addlegendentry{$\bm{-\phi_1(x)}$}

                    \addplot[solid,{black}, line width=1 pt, mark = none] table[x index=0, y index=1] {P4_buck_mode_phi_2_minus_phi_1_LBFGSB_method.txt};
                    \addlegendentry{$\bm{\phi_2(x) - \phi_1(x)}$}   
				\end{axis}
			\end{tikzpicture}
			\caption{}
        \label{Fig:P4_buck_mode_LBFGSB_method}
		\end{subfigure}
		\caption{Eigenvectors $\phi_1(x)$ and $\phi_2(x)$ correspond to the lateral nodal displacements of a 1,000-element 1D column at the optimum point, obtained using: (a) the conventional aggregation function ($\rho=500$), and (b) the proposed method. All the modes are normalized such that their maximum deformation is equal to 1.} 
        \label{P4_buck_modes}
	\end{figure}
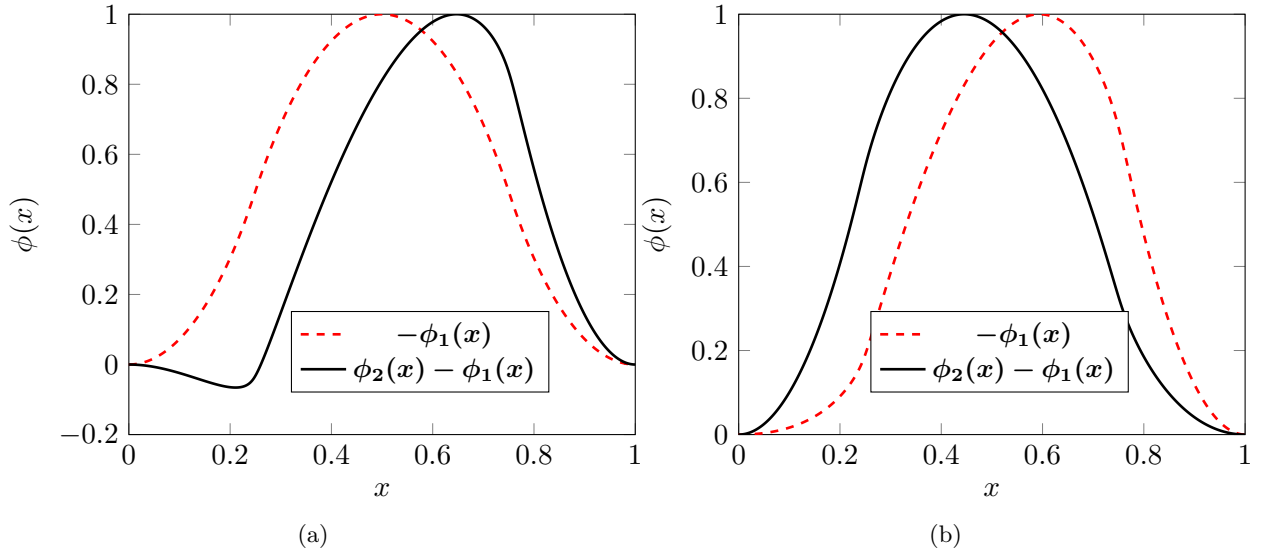


\begin{figure}
		\centering
		\begin{subfigure}[c]{0.44\textwidth}
			\centering
			\includegraphics[width=\textwidth]{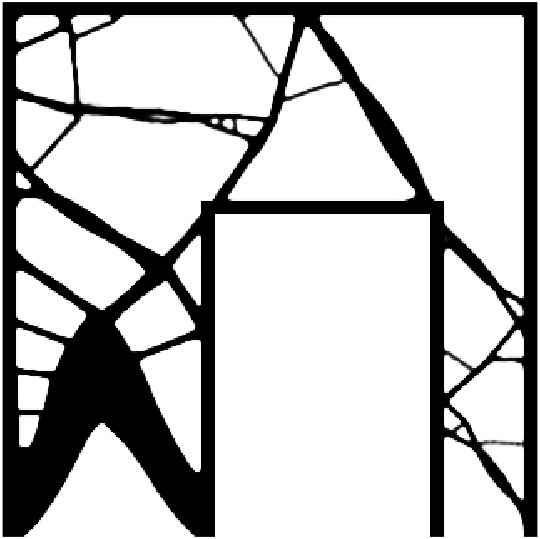} 
			\caption{}
			\label{Fig:P2_LBFGSB_method_design}
		\end{subfigure}
		\hfill
		\begin{subfigure}[c]{0.44\textwidth}
			\centering
			\includegraphics[width=4cm,height = 9.3cm]{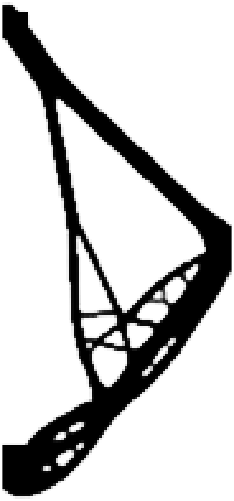} 
			\caption{}
			\label{Fig:P3_LBFGSB_method_design}
		\end{subfigure}
		\caption{The optimized design obtained after minimizing the difference between the eigenvalues (a) Wall reinforcement problem (b) Shear-loaded hinged plate}
        \label{Fig:P2_and_P3_LBFGSB_designs}
\end{figure}
Optimality condition for Eq.~\ref{Eq:1Dcolumcolopt} is written as:
   
\begin{equation} \label{P4_LBFGSB_gradient}
       \left( \sum_{i=1}^{n_e - 1}\sum_{j= i+1}^{n_e} 2\left(\frac{r_i - r_j}{r_0^2}\right) (\partial r_i - \partial r_j)\right)+ \left(2g_V(\mathbf{A})(\nabla g_V)\right) = 0
   \end{equation}

 Note that, although the initial guess is already close to the final solution, the volume constraint is also included in the unconstrained objective function. This is because changing $V_0$ in $g_V(\mathbf{A})$ causes the first and second eigenvalues to coalesce at different values. This method helps identify the modality via the optimality condition for the modified optimization problem (Eq.~\ref{P4_LBFGSB_gradient}). Near the initial guess, typically $\partial r_i \neq \partial r_j$ and $\nabla g_V \neq 0$. To satisfy Eq.~\ref{P4_LBFGSB_gradient}, the algorithm drives the solution toward a point where $r_i = r_j$ and $g_V = 0$. Consequently, the modified problem seeks a feasible point where $n_e$ eigenvalues coalesce. However, this point is not necessarily the true optimum of the original problem. We explore this concept in more detail when analyzing the wall reinforcement problem.


Fig.~\ref{P4_buck_modes} illustrates the importance of correctly determining the optimal modality, showing the lateral displacement eigenvectors $\phi_1(x)$ and $\phi_2(x)$ for the discretized column (Fig.~\ref{Fig:P4_Schematic}). Fig.~\ref{Fig:P4_buck_mode_conven_method} shows two linear combinations of these eigenvectors at the optimum obtained via the conventional aggregation method: the mode $-\phi_1(x)$ is symmetric about the column center, whereas the second mode is asymmetric with negative deflection near the end. In contrast, the proposed optimization formulation yields fundamentally different mode shapes (Fig.~\ref{Fig:P4_buck_mode_LBFGSB_method}). Here, both modes are asymmetric about the center and exhibit strictly positive deflection, closely matching the buckled shapes reported in~\cite{olhoff1977single}. This demonstrates that the conventional aggregation method produces a unimodal solution, rendering a linear combination of its eigenvectors invalid. The proposed method, however, yields the correct solution.

Having demonstrated the proposed approach on the 1D compressed column, we now apply it to the fake coalescence case shown in Fig.~\ref{Fig:Fake coalescence}. Using the conventional aggregation method with $\rho = 100$, the optimum point is $x = -0.0102$, where the relative eigenvalue difference is $1.5201 \times 10^{-4}$.  Applying the proposed strategy to the same problem yields an optimal solution at $x = 0$, with a relative difference of exactly $10^{-4}$.  The relative difference between the eigenvalues changes very little between the two methods, which signals fake coalescence. When eigenvalues do not truly coalesce, the standard aggregation method already produces a solution very close to the true optimum. In this way, the proposed strategy helps identify fake coalescence and remove irrelevant eigenmodes at the optimum point.

\subsubsection{Modality of reinforced wall problem} \label{sec:modreiforcedwall}
\begin{figure}[h!]
		\centering
		\begin{subfigure}{0.3\textwidth}
			\centering
			\includegraphics[width=\linewidth]{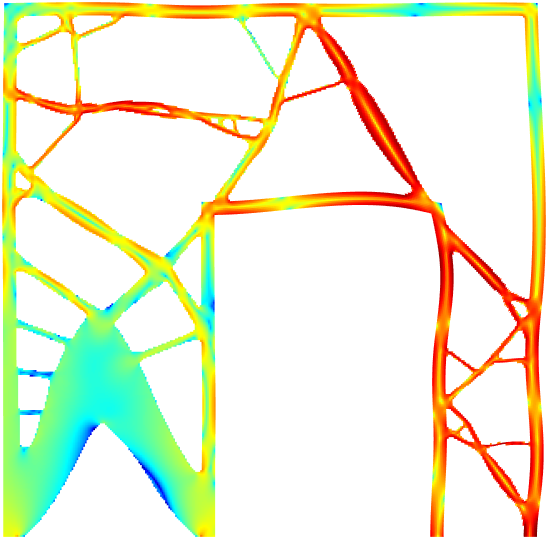}
			\caption{}
			\label{Fig:P2_buck_mode_actual_phi_1}
		\end{subfigure}
		\hfill
		\begin{subfigure}{0.3\textwidth}
			\centering
			\includegraphics[width=\linewidth]{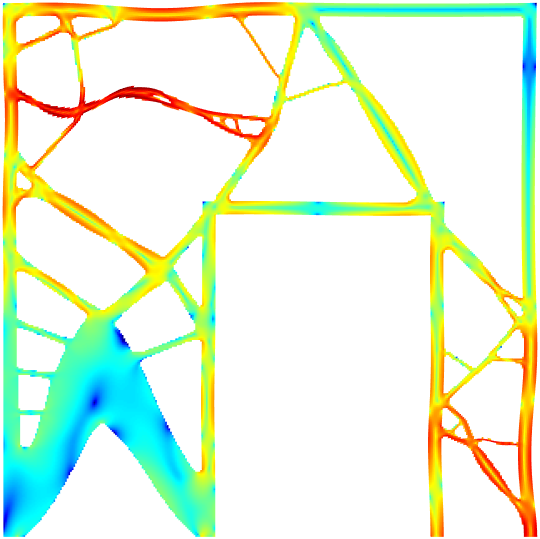}
			\caption{}
			\label{Fig:P2_buck_mode_actual_phi_2}
		\end{subfigure}
		\hfill
		\begin{subfigure}{0.3\textwidth}
			\centering
			\includegraphics[width=\linewidth]{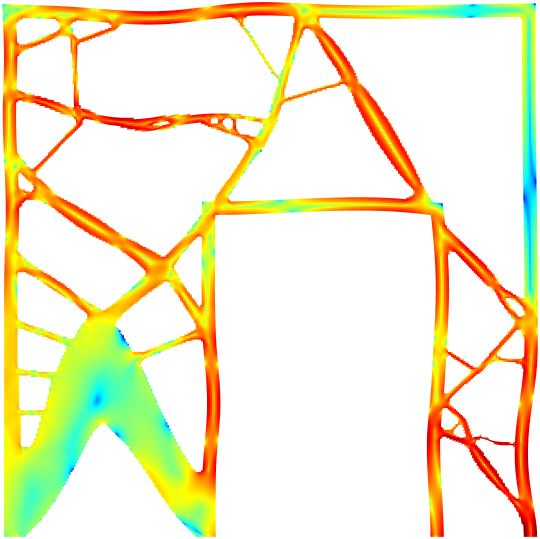}
			\caption{}
			\label{Fig:P2_buck_mode_actual_phi_3}
		\end{subfigure}
		
		\vspace{0.5cm}
		
		\begin{subfigure}{0.3\textwidth}
			\centering
			\includegraphics[width=\linewidth]{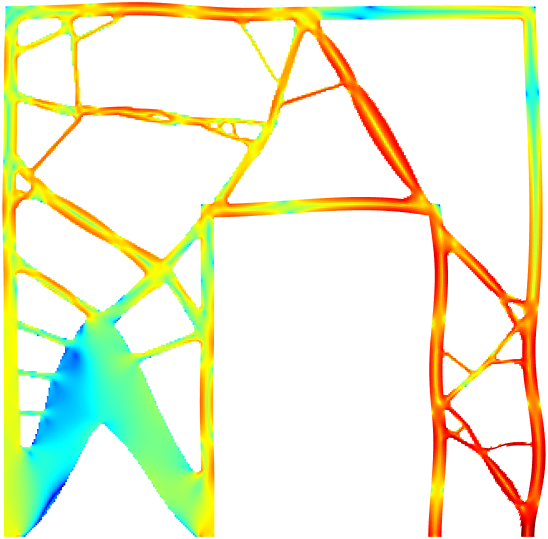}
			\caption{}
			\label{Fig:P2_buck_mode_LC_phi_1}
		\end{subfigure}
		\hfill
		\begin{subfigure}{0.3\textwidth}
			\centering
			\includegraphics[width=\linewidth]{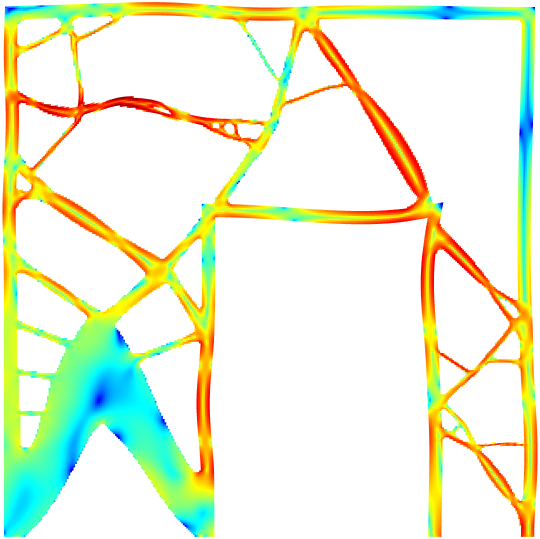}
			\caption{}
			\label{Fig:P2_buck_mode_LC_phi_2}
		\end{subfigure}
		\hfill
		\begin{subfigure}{0.3\textwidth}
			\centering
			\includegraphics[width=\linewidth]{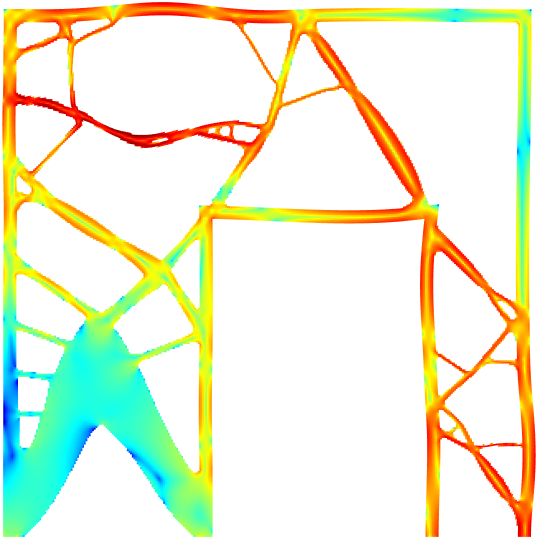}
			\caption{}
			\label{Fig:P2_buck_mode_LC_phi_3}
		\end{subfigure}
		
		\caption{Buckled mode shapes for the wall reinforcement problem corresponding to the final iteration of the new strategy (a) First Mode shape ($\phi_1$), (b) Second Mode shape ($\phi_2$), (c) Third Mode shape ($\phi_3$). Eigenmodes obtained by the linear combination of the first 3 eigenvectors (without considering fixed dofs) 
			(d) $\tilde{\phi_1} = (6\phi_1 - 2\phi_2 - 3\phi_3)/7$ 
			(e) $\tilde{\phi_2} = (-2\phi_1 + 3\phi_2 - 6\phi_3)/7$
			(f) $\tilde{\phi_3} = (-3\phi_1 - 6\phi_2 - 2\phi_3)/7$. The colormap (blue to red) shows the logarithm of normalized strain energy density $log\left(\frac{s_e}{s_{e,max}}\right)$ (low to high).}
		\label{Fig:P2_buck_mode_LBFGSB_method}
	\end{figure}

We now determine the modality for the reinforced wall problem (Fig.~\ref{Fig:P2_Schematic}), where five eigenvalues are nearly equal. Because these eigenvalues act as constraints, we also include the buckling constraint in the objective function while minimizing the differences between them. The unconstrained optimization problem is formulated as follows:
	\begin{equation}
    \begin{aligned}
		& \min_{\bm{x}} \quad \left( \sum_{i=1}^{n_e}\sum_{j= i+1}^{n_e} \left(\frac{r_i - r_j}{\overline{r}}\right)^2 \right) + \left(g_C(\mathbf{x})\right)^2 +  \left(\frac{\sum_{i=1}^{n_e} \hspace{0.1em} r_{i}(\bm{x})}{n_e\overline{r}} - 1\right)^2 \\  
    \end{aligned}
    \label{minimization_formulation_prob_2}
	\end{equation}

where $\overline{r} = 1/\overline{\lambda}$. Here, we use only the first symmetric polynomial, which is the sum of the eigenvalues, to model the buckling constraint. This term is sufficient because minimizing the differences between eigenvalues already ensures that $r_i = \overline{r}$. We start with $n_e = 5$ and gradually reduce its value until the relative difference between the eigenvalues is very small compared to that of our initial guess. The L-BFGS-B algorithm~\cite{byrd1995limited,zhu1997algorithm,morales2011remark} is used as per the tolerance and other parameters defined earlier. We consider different $n_e$ for solving this problem, and summarize the results in Table~\ref{Tab:P2_results_LBFGSB_method}.  When $n_e = 3$, the objective function obtained using the L-BFGS-B algorithm ($f(x)/f_0 = 0.341479$) is very close to that obtained from the ELTD method ($f(x)/f_0 = 0.34153$). We achieve a relative difference of $(r_1 - r_2)/r_1 = 4.4857 \times 10^{-13}$ and $(r_1 - r_3)/r_1 = 7.5127 \times 10^{-13}$ in 62 outer iterations. Therefore, the modality at the optimum point for the wall reinforcement problem (Fig.~\ref{Fig:P2_Schematic}) with $\overline{\lambda} = 0.3$ is 3. 

\begin{table}[H]
\centering
\caption{Results for the wall reinforcement problem across different values of $n_e$, obtained by solving Eq.~\ref{minimization_formulation_prob_2}. Here, $g_{\lambda_1}(x) = (\lambda_1/\overline{\lambda}) - 1$, $L_{\infty}\text{ norm} = \max(\vert{}\text{xPhys}_{\text{IG}} - \text{xPhys}_{\text{final}}\vert{})$, and $\% \text{ change in design} = \frac{\Vert{}\text{xPhys}_{\text{IG}} - \text{xPhys}_{\text{final}}\Vert{}_2}{\Vert{}\text{xPhys}_{\text{IG}}\Vert{}_2} \times 100$.}
 \label{Tab:P2_results_LBFGSB_method}
\renewcommand{\arraystretch}{1.15}
\setlength{\tabcolsep}{4pt}

\resizebox{\textwidth}{!}{%
\begin{tabular}{c| c| c| c| c| c| c| c}
\hline\hline
$n_e$ &
$f/f_0$ &
$g_{\lambda_1}(x)$ &
$g_C(x)$ &
Eigenvalues &
Iterations &
$L_{\infty}$ norm &
\% change in design
\\
\hline

2 &
0.341397 &
$-6.07\times10^{-13}$ &
$1.87\times10^{-14}$ &
\begin{tabular}{c}
0.3000, 0.3000, 0.3023,\\
0.3030, 0.3075
\end{tabular} &
22 &
0.0082 &
0.1278
\\

3 &
0.341479 &
$-6.93\times10^{-13}$ &
$-5.76\times10^{-13}$ &
\begin{tabular}{c}
0.3000, 0.3000, 0.3000,\\
0.3025, 0.3074
\end{tabular} &
62 &
0.2037 &
0.7880
\\

4 &
0.342741 &
$-3.56\times10^{-9}$ &
$2.45\times10^{-9}$ &
\begin{tabular}{c}
0.3000, 0.3000, 0.3000,\\
0.3000, 0.3025
\end{tabular} &
697 &
0.5861 &
3.9376
\\

5 &
0.342630 &
$-8.71\times10^{-4}$ &
$3.22\times10^{-4}$ &
\begin{tabular}{c}
0.2997, 0.2999, 0.3001,\\
0.3003, 0.3005
\end{tabular} &
152 &
0.6533 &
3.5143
\\

\hline\hline
\end{tabular}%
}
\end{table}

Table~\ref{Tab:P2_results_LBFGSB_method} suggests that the optimal solution could be bimodal, trimodal, or tetramodal. Sorting these cases by objective value might initially point to a bimodal solution ($n_e = 2$). However, the problem is not bimodal, even though the objective value for $n_e = 2$ is slightly lower than for $n_e = 3$. This slight difference arises because the compliance constraint is overshot, thereby artificially reducing the volume fraction. Likewise, the problem cannot be tetramodal because its objective value is significantly higher, despite having higher compliance and a lower buckling load. Additionally, the design changes substantially from the initial guess. Therefore, the wall reinforcement problem is trimodal for these parameters, as confirmed by the final optimized design in Fig.~\ref{Fig:P2_LBFGSB_method_design}, which exhibits the same structural features as Fig.~\ref{Fig:P2_ELTD_method_design}.

The top three panels in Fig.~\ref{Fig:P2_buck_mode_LBFGSB_method} show the eigenvectors from the final iteration of the proposed strategy. The bottom three panels show the mode shapes obtained by multiplying these eigenvectors by an orthogonal matrix. As a result, the top and bottom mode shapes differ distinctly in both deformation patterns and strain energy density. Without first identifying the optimal modality, generating and visualizing such a wide variety of feasible mode shapes would not be possible.

%

\begin{table}[H]
\centering
\caption{Results for the shear-loaded hinged plate problem across different values of $n_e$, obtained by solving Eq.~\ref{minimization_formulation_prob_3}. Here, $g_{\lambda_1}(x) = (\lambda_1/\overline{\lambda}) - 1$, $L_{\infty}\text{ norm} = \max(\vert{}\text{xPhys}_{\text{IG}} - \text{xPhys}_{\text{final}}\vert{})$, and $\% \text{ change in design} = \frac{\Vert{}\text{xPhys}_{\text{IG}} - \text{xPhys}_{\text{final}}\Vert{}_2}{\Vert{}\text{xPhys}_{\text{IG}}\Vert{}_2} \times 100$.}
\label{Tab:P3_results_LBFGSB_method}
\renewcommand{\arraystretch}{1.15}
\setlength{\tabcolsep}{4pt}

\resizebox{\textwidth}{!}{%
\begin{tabular}{c| c| c| c| c| c| c| c}
\hline\hline
$n_e$ &
$f/f_0$ &
$g_{\lambda_1}(x)$ &
$g_V(x)$ &
Eigenvalues &
Iterations &
$L_{\infty}$ norm &
\% change in design
\\
\hline

2 &
1.585731 &
$-2.92\times10^{-13}$ &
$1.14\times10^{-13}$ &
\begin{tabular}{c}
0.7500, 0.7500, 0.7674, 0.7836,\\
0.8048, 0.8156, 0.8440
\end{tabular} &
70 &
0.0404 &
0.1634
\\

3 &
1.586875 &
$-1.05\times10^{-13}$ &
$-1.24\times10^{-13}$ &
\begin{tabular}{c}
0.7500, 0.7500, 0.7500, 0.7641,\\
0.7863, 0.8035, 0.8366
\end{tabular} &
128 &
0.0491 &
0.3541
\\

4 &
1.586148 &
$1.88\times10^{-11}$ &
$1.11\times10^{-12}$ &
\begin{tabular}{c}
0.7500, 0.7500, 0.7500, 0.7500,\\
0.7681, 0.7990, 0.8421
\end{tabular} &
475 &
0.0915 &
0.4304
\\

5 &
1.592239 &
$-2.25\times10^{-4}$ &
$2.15\times10^{-5}$ &
\begin{tabular}{c}
0.7498, 0.7499, 0.7499, 0.7499,\\
0.7499, 0.8021, 0.8225
\end{tabular} &
406 &
0.2277 &
1.4397
\\

6 &
1.600100 &
$-8.85\times10^{-4}$ &
$4.30\times10^{-4}$ &
\begin{tabular}{c}
0.7493, 0.7494, 0.7495, 0.7495,\\
0.7495, 0.7497, 0.8033
\end{tabular} &
731 &
0.6273 &
3.7423
\\

7 &
1.596461 &
$-2.32\times10^{-3}$ &
$1.38\times10^{-3}$ &
\begin{tabular}{c}
0.7483, 0.7485, 0.7486, 0.7488,\\
0.7489, 0.7492, 0.7495
\end{tabular} &
503 &
0.5974 &
3.5736
\\

\hline\hline
\end{tabular}%
}
\end{table}

    \begin{figure}[h!]
		\centering
		\begin{subfigure}[c]{0.23\textwidth}
			\centering
			\includegraphics[width=0.6\textwidth,height = 8cm]{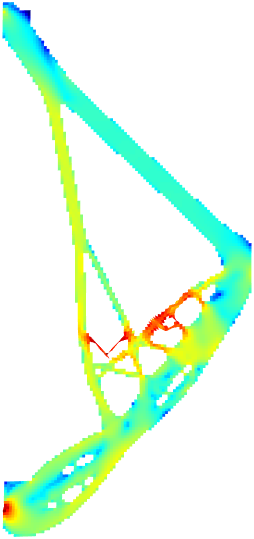} 
			\caption{}
			\label{Fig:P3_buck_mode_actual_phi_1}
		\end{subfigure}
		\begin{subfigure}[c]{0.23\textwidth}
			\centering
			\includegraphics[width=0.6\textwidth,height = 8cm]{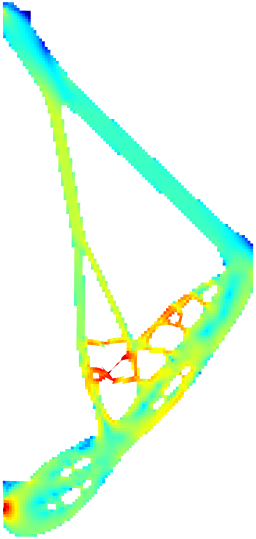} 
			\caption{}
			\label{Fig:P3_buck_mode_actual_phi_2}
		\end{subfigure}
        \begin{subfigure}[c]{0.23\textwidth}
			\centering
			\includegraphics[width=0.6\textwidth,height = 8cm]{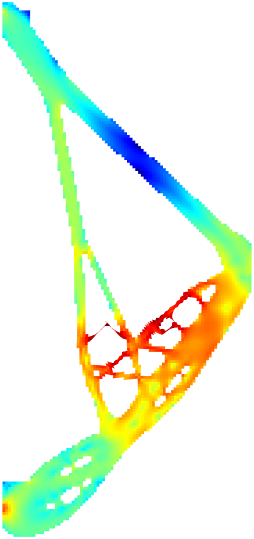} 
			\caption{}
			\label{Fig:P3_buck_mode_actual_phi_3}
		\end{subfigure}
        \begin{subfigure}[c]{0.23\textwidth}
			\centering
			\includegraphics[width=0.6\textwidth,height = 8cm]{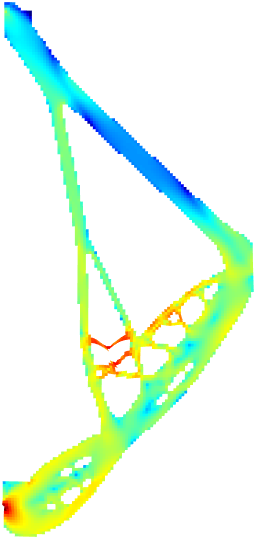} 
			\caption{}
			\label{Fig:P3_buck_mode_actual_phi_4}
		\end{subfigure}

        \vspace{0.5cm}

        	\begin{subfigure}[c]{0.23\textwidth}
			\centering
			\includegraphics[width=0.6\textwidth,height = 8cm]{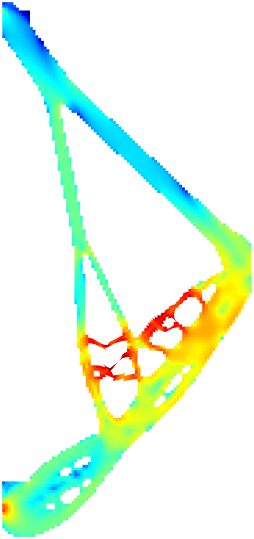} 
			\caption{}
			\label{Fig:P3_buck_mode_LC_phi_1}
		\end{subfigure}
		\begin{subfigure}[c]{0.23\textwidth}
			\centering
			\includegraphics[width=0.6\textwidth,height = 8cm]{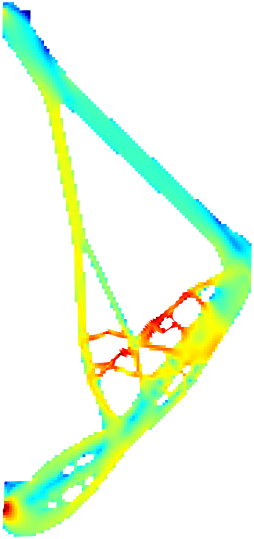} 
			\caption{}
			\label{Fig:P3_buck_mode_LC_phi_2}
		\end{subfigure}
        \begin{subfigure}[c]{0.23\textwidth}
			\centering
			\includegraphics[width=0.6\textwidth,height = 8cm]{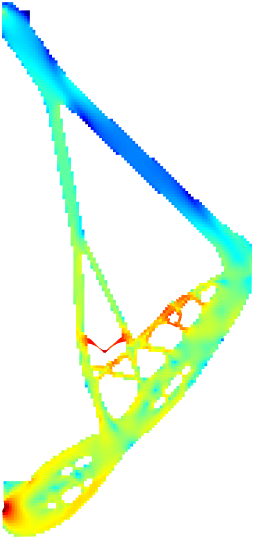} 
			\caption{}
			\label{Fig:P3_buck_mode_LC_phi_3}
		\end{subfigure}
        \begin{subfigure}[c]{0.23\textwidth}
			\centering
			\includegraphics[width=0.6\textwidth,height = 8cm]{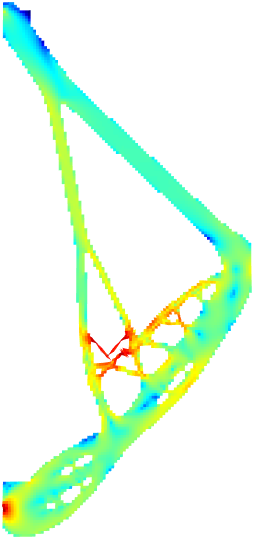} 
			\caption{}
			\label{Fig:P3_buck_mode_LC_phi_4}
		\end{subfigure}
        
\caption{Buckled mode shapes for the shear-loaded hinged plate at the final iteration of the proposed strategy: (a) first mode shape ($\phi_1$), (b) second mode shape ($\phi_2$), (c) third mode shape ($\phi_3$), and (d) fourth mode shape ($\phi_4$). Mode shapes obtained from linear combinations of the first four eigenvectors, excluding fixed degrees of freedom:
(e) $\tilde{\phi}_1 = (\phi_1 + \phi_2 + \phi_3 + \phi_4)/2$,
(f) $\tilde{\phi}_2 = (\phi_1 - \phi_2 + \phi_3 - \phi_4)/2$,
(g) $\tilde{\phi}_3 = (\phi_1 + \phi_2 - \phi_3 - \phi_4)/2$, and
(h) $\tilde{\phi}_4 = (\phi_1 - \phi_2 - \phi_3 + \phi_4)/2$.
The color bar from blue to red indicates the logarithm of the normalized strain energy density, $\log\left(\frac{s_e}{s_{e,\text{max}}}\right)$, ranging from low to high.}
		\label{Fig:P3_buck_mode_LBFGSB_method}
\end{figure}

\subsubsection{Modality of shear-loaded hinged plate problem}\label{sec:modshearplate}
Herein, the modality of the shear-loaded hinge plate problem is presented. The formulated constrained optimization setting is similar to that of the reinforced wall problem, which is given as: 
 \begin{equation}
 \begin{aligned}
		& \min_{\bm{x}} \quad \left( \sum_{i=1}^{n_e}\sum_{j= i+1}^{n_e} \left(\frac{r_i - r_j}{\overline{r}}\right)^2 \right) + \left(g_V(\mathbf{x})\right)^2 +  \left(\frac{\sum_{i=1}^{n_e} \hspace{0.1em} r_{i}(\bm{x})}{n_e\overline{r}} - 1\right)^2 \\   
\end{aligned}
\label{minimization_formulation_prob_3}
\end{equation}
 The L-BFGS-B algorithm~\cite{byrd1995limited,zhu1997algorithm,morales2011remark}  with the tolerance and other parameters defined earlier is used to solve the above optimization problem. We consider different values of $n_e$ starting from $n_e = 7$, and summarize the results in Table~\ref{Tab:P3_results_LBFGSB_method}. The objective function value obtained from the ELTD method is $f(x)/f_0 = 1.5858$. This table shows that the optimal solution could be bimodal, trimodal, or tetramodal ($n_e = 2, 3,$ or $4$). Sorting these cases by objective value ranks $n_e = 2$ lowest, followed by $n_e = 4$ and $n_e = 3$. However, the solution is not bimodal because a slightly positive volume constraint artificially reduces the compliance relative to the ELTD method. The results for $n_e = 3$ and $n_e = 4$ are very close. Because their constraint values have opposite signs, the exact optimum likely lies between them, closer to $n_e = 4$, which has a significantly lower objective value. Thus, the shear-loaded hinged plate problem is tetramodal for this parameter set. The method achieves relative differences of $(r_1 - r_2)/r_1 = 6.0458 \times 10^{-12}$, $(r_1 - r_3)/r_1 = 1.5428 \times 10^{-11}$, and $(r_1 - r_4)/r_1 = 2.7772 \times 10^{-11}$. The optimized design remains very close to the initial design (Fig.~\ref{Fig:P3_LBFGSB_method_design}). The associated buckled modes, along with a representative linear combination, are shown in Fig.~\ref{Fig:P3_buck_mode_LBFGSB_method}.


\begin{figure}
		\begin{subfigure}[t]{0.48\textwidth}
        \centering
			\begin{tikzpicture}
				\begin{axis}[
					xlabel={Iteration},
					ylabel={Objective function},
					xmin=1,xmax=70,
					ymin=-5, ymax=100,
                    width=\linewidth,
                    ]
					\addplot[solid,{black}, line width=1 pt] table[x index=0, y index=1] {P2_Obj_val_variation_LBFGSB_method.txt}; 
				\end{axis}
			\end{tikzpicture}
			\caption{}
       \label{Fig:P2_Obj_val_variation_LBFGSB_method}
		\end{subfigure}
		\hfill
		\begin{subfigure}[t]{0.48\textwidth}
        \centering
			\begin{tikzpicture}
				\begin{axis}[
					xlabel={Iteration},
					ylabel={Objective function},
					xmin=1,xmax=500,
					ymin=-1, ymax= 20,
                    width=\linewidth,
                    ]
					\addplot[solid,{black}, line width=1 pt] table[x index=0, y index=1] {P3_Obj_val_variation_LBFGSB_method.txt}; 
				\end{axis}
			\end{tikzpicture}
			\caption{}
       \label{Fig:P3_Obj_val_variation_LBFGSB_method}
		\end{subfigure}
		\caption{Convergence history of the objective function for finding the optimal modality using the proposed method: (a) wall reinforcement problem and (b) shear loaded hinged plate.} 
      \label{Fig:Convergence_plots_for_LBFGSB_method}
	\end{figure}
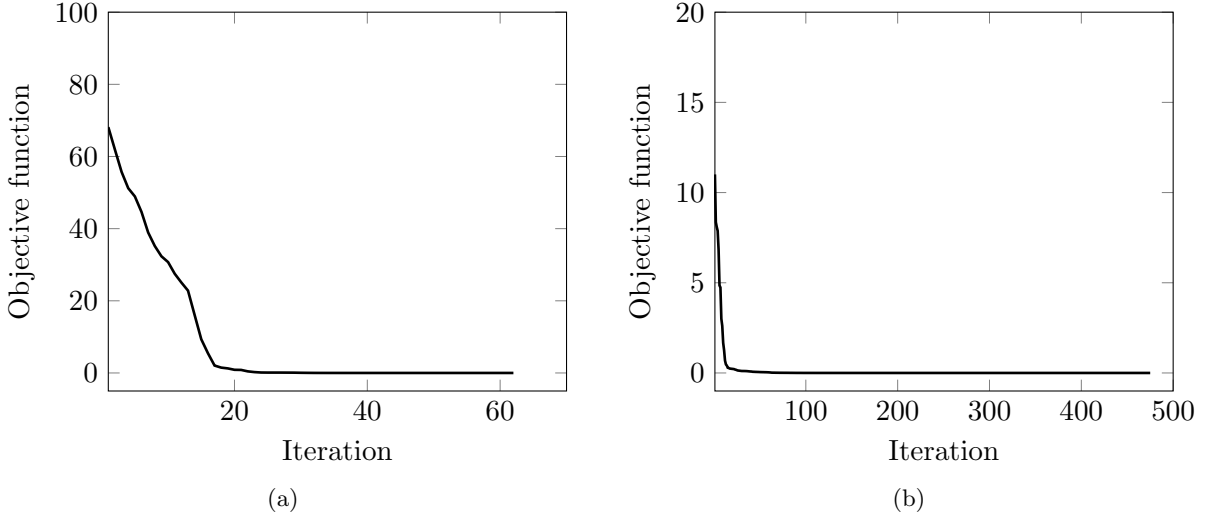
   Figure~\ref{Fig:Convergence_plots_for_LBFGSB_method} shows the objective function convergence history for both the wall reinforcement (Sec.~\ref{sec:modreiforcedwall}) and shear loaded hinged plate (Sec.~\ref{sec:modshearplate}) problems. Both plots display a steady decrease in the objective function without any fluctuations, demonstrating the robustness of the proposed approach. Using this same routine, we have already confirmed that the compressed column described in Sec.~\ref{sec:mod1Dcompressed} is unimodal.

\section{Closure} \label{Sec5:Conc}
This paper addresses two major challenges in structural optimization using aggregation functions: selecting the appropriate number of eigenvalues during optimization and accurately determining the modality at the optimal point.

We show that choosing an arbitrary number of eigenvalues throughout optimization can produce incorrect gradients when higher-order eigenvalues coalesce. To overcome this, we propose the ELTD scheme, which selects only the necessary eigenvalues at each iteration based on their relative differences. This approach significantly reduces computational time, achieving nearly a 50 percent reduction in the unimodal 2D compressed column problem. The robustness of the ELTD scheme is further demonstrated in the wall-reinforcement and shear-loaded hinged-plate problems, where it offers substantial performance gains when dealing with closely clustered eigenvalues.

Accurately determining modality is essential for visualizing all valid buckled mode shapes. Conventional aggregation functions often fail to detect true eigenvalue coalescence, rendering linear combinations of their eigenvectors invalid. Furthermore, relying solely on relative differences can create false coalescence. To resolve these issues, we introduce a new strategy that formulates an optimization problem to directly minimize the differences between eigenvalues. This strategy achieves precise results with relative differences on the order of $10^{-10}$ in very few iterations while reliably identifying false coalescence.

Using the proposed strategy with the L-BFGS-B algorithm, we successfully identify the modal behavior of various structural problems. Importantly, our approach provides the correct bimodal modality for the 1D compressed column problem as established in the literature, while showing that the wall reinforcement and shear-loaded hinged plate problems are trimodal and tetramodal, respectively. The resulting buckled mode shapes and their linear combinations display clear variations in deformation patterns and strain energy density across modes. Ultimately, this method provides a dependable framework for identifying optimal modalities and visualizing complete buckling behaviors in complex structures.

\appendix

\appendix
\section{Value of $\epsilon$ in ELTD scheme}\label{app:app1}

Equation~\ref{Eq:epsilon-ln} outlines the ELTD scheme for choosing which eigenvalues to include in aggregation during each optimization step. If the difference between $r_1$ and $r_i$ exceeds $\left(-\ln(\epsilon)/\rho \right)$, those eigenvalues are excluded because any exponential weight $w_i$ smaller than $\epsilon$ contributes negligibly to the sum. The choice of $\epsilon$ ($10^{-9}$ in this work) depends on two main factors: the weight of the exponential term and the need to avoid abrupt gradient changes near multimodal points. To evaluate the effect of $\epsilon$, a parametric study is conducted on the 1D clamped column problem (Fig.~\ref{Fig:P4_Schematic}) across various $\epsilon$ values.

We solve this problem using different $\epsilon$ values for aggregation parameters of 100 and 500 until the $L_{\infty}$ norm of design variable changes drops below $10^{-4}$. The final optimal eigenvalues remain nearly identical across all $\epsilon$ values. However, smaller $\epsilon$ values detect the onset of bimodality earlier, which helps reduce oscillations caused by frequent switching between unimodal and bimodal states. While $\epsilon = 10^{-3}$ is sufficient for an aggregation parameter of 100, an aggregation parameter of 500 requires a smaller value of $10^{-6}$. Higher aggregation parameters make the function sharper near multimodal points, leading to rapid gradient changes that require smaller thresholds for smooth transitions. Taking a conservative approach for higher dimensional problems, we recommend setting $\epsilon = 10^{-9}$ for aggregation parameters up to at least 1000. In practice, most buckling optimization problems use aggregation parameters below 200 due to nonlinearity and convergence challenges.

\begin{table}[H]
\centering
\caption{Effect of $\epsilon$ on the optimal solution and optimization history of 1D column problem for different values of the aggregation parameter $\rho$.}
\label{Tab:P4_parametric_study_on_eps_values}
\renewcommand{\arraystretch}{1.15}
\setlength{\tabcolsep}{4pt}

\resizebox{\textwidth}{!}{%
\begin{tabular}{c| c| c| c |c |c |c |c}
\hline\hline
$\rho$ &
$\epsilon$ &
$-\ln(\epsilon)/\rho$ &
First iteration with $\lambda_2$ &
No. of modality swaps &
$\lambda_1$ &
$\lambda_2$ &
No. of eigenvalues aggregated
\\
\hline

\multirow{9}{*}{100}
& $10^{-1}$ & 0.0230 & 14 & 55 & 4.3339 & 4.8067 & 2 \\
& $10^{-2}$ & 0.0461 & 13 & 1  & 4.3339 & 4.8068 & 2 \\
& $10^{-3}$ & 0.0691 & 11 & 1  & 4.3339 & 4.8069 & 2 \\
& $10^{-4}$ & 0.0921 & 9  & 1  & 4.3339 & 4.8067 & 2 \\
& $10^{-5}$ & 0.1151 & 7  & 1  & 4.3338 & 4.8070 & 2 \\
& $10^{-6}$ & 0.1382 & 4  & 1  & 4.3339 & 4.8066 & 3 \\
& $10^{-7}$ & 0.1612 & 1  & 0  & 4.3339 & 4.8069 & 3 \\
& $10^{-8}$ & 0.1842 & 1  & 0  & 4.3338 & 4.8072 & 4 \\
& $10^{-9}$ & 0.2072 & 1  & 0  & 4.3338 & 4.8071 & 7 \\
\hline

\multirow{11}{*}{500}
& $10^{-1}$  & 0.0046 & 16 & 377 & 4.3570 & 4.4454 & 2 \\
& $10^{-2}$  & 0.0092 & 16 & 23  & 4.3574 & 4.4673 & 2 \\
& $10^{-3}$  & 0.0138 & 15 & 9   & 4.3575 & 4.4676 & 2 \\
& $10^{-4}$  & 0.0184 & 15 & 7   & 4.3575 & 4.4674 & 2 \\
& $10^{-5}$  & 0.0230 & 14 & 3   & 4.3574 & 4.4673 & 2 \\
& $10^{-6}$  & 0.0276 & 14 & 3   & 4.3574 & 4.4673 & 2 \\
& $10^{-7}$  & 0.0322 & 14 & 1   & 4.3575 & 4.4672 & 2 \\
& $10^{-8}$  & 0.0368 & 13 & 1   & 4.3574 & 4.4674 & 2 \\
& $10^{-9}$  & 0.0414 & 13 & 1   & 4.3574 & 4.4674 & 2 \\
& $10^{-12}$ & 0.0553 & 12 & 1   & 4.3575 & 4.4674 & 2 \\
& $10^{-15}$ & 0.0691 & 11 & 1   & 4.3574 & 4.4672 & 2 \\

\hline\hline
\end{tabular}%
}
\end{table}

\end{document}